\documentclass[preprint2]{aastex}

\slugcomment{Accepted for publication in the ApJ; submitted 2022 August 25, accepted September 16.}

\shorttitle{Variability in PPNe: IX. Evolution in a Decade}

\shortauthors{Hrivnak et al.}

\begin{document}

\title{Variability in Protoplanetary Nebulae: IX. Evidence for Evolution in a Decade} 


\author{Bruce J. Hrivnak\altaffilmark{1},  Wenxian Lu\altaffilmark{1}, William C. Bakke\altaffilmark{1}, and Peyton J. Grimm\altaffilmark{1}}

\altaffiltext{1}{Department of Physics and Astronomy, Valparaiso University, 
Valparaiso, IN 46383, USA; bruce.hrivnak@valpo.edu, wen.lu@valpo.edu (retired), william.bakke.valpo.edu, peyton.grimm@valpo.edu}

\begin{abstract}
We have carried out a new photometric {\it V,R}$_C$ study of 12 protoplanetary nebulae, objects in the short-lived transition between the asymptotic giant branch and planetary nebula phases of stellar evolution.
These had been the subjects of an earlier study, using data from 1994$-$2007, that found that all 12 varied periodically, with pulsation periods in the range of $\sim$38 to $\sim$150 days.
They are all carbon-rich, with F$-$G spectral types.
We combined our new (2008$-$2018) data with publicly available ASAS-SN data and determined new periods for their variability.
The older and newer period values were compared to investigate evidence of period change, for which there is theoretical support that it might be detectable in a decade or two in some cases.
Such a detection is challenging since the light curves are complicated, with multiple periods, changing amplitudes, and evidence of shocks.
Nevertheless, we found one, and possibly two, such cases, which are associated with the higher-temperature stars in the sample (7250 and 8000 K).  
These results are most consistent with the evolution of stars at the lower end of the mass range of carbon stars, $\sim$1.5$-$2 M$_{\sun}$.
Several of the stars show longer-term trends of increasing (six cases) or decreasing (one case) brightness, which we think most likely due to changes in the circumstellar dust opacity.
There is one case of a possible $\sim$1.8 yr period in addition to the shorter pulsation.  This is interpreted as possible evidence of an orbiting companion.

\end{abstract}

\section{Introduction}

Protoplanetary (or preplanetary) nebulae (PPNe) are intermediate- and low-mass stars (1$-$8 M$_{\sun}$) in the short-lived (several thousand years) transition from the asymptotic giant branch (AGB) to the planetary nebula (PN) phases of stellar evolution.  They have lost almost all of their atmosphere during the red giant and AGB phases and are surrounded by an expanding circumstellar envelope of gas and dust.  As such, their spectral energy distributions (SEDs) show a combination of a reddened photosphere, peaking in the visible/near-infrared, and the mid-infrared emission from the surrounding dust, giving them a characteristic double-peaked appearance \citep{partha86,hri89}.  Other characteristics include molecular emissions (CO, OH, HCN) at millimeter and radio wavelengths \citep{lik89,lik91,omont93} and silicate or carbon-based mid-infrared dust emission features \citep{mol02,hri09}. 

In our initial study of the light variability of PPNe, we investigated 12 carbon-rich (C-rich) objects, using observations obtained at the Valparaiso University Observatory (VUO) from 1994$-$2007. 
Their C-rich designation was based on the presence of C$_2$ and C$_3$ absorption features in their visible spectra \citep{hri95}, abundance analyses based on high-resolution visible-band spectra \citep{vanwin00,red99,red02}, and the presence of C-rich features in their mid-infrared spectra \citep{hri09}. 
Periodic light variations were found in all 12, with periods ranging from $\sim$35 to $\sim$150 days \citep{hri10}, although some results are more secure than others.  These light variations are attributed to pulsation.  
We have carried out subsequent studies of four of these \citep{hri13,hri20a} and six additional southern hemisphere C-rich sources \citep{hri20a,hri21}.

Monitoring of the light curves of 11 of these objects has been continued over 10 more seasons (2008$-$2018), with better precision.
During this time, all-sky survey data also became publicly available, and some of these data are included in this study.  We also wanted to extend the temporal baseline to see if any clear changes in period might be discerned, as that was one of the predictions of the initial study.  Another goal was to investigate if longer periods are present in the data, which could possibly indicate a binary nature.   

In this paper, we report on an updated study of the initial 12 sources and then present a comparative discussion of the results of the period analyses of all of the studied C-rich sources.  We begin with a description of our newer observations and the additional data sets used, followed by  results of the period analyses. 
These new period values are compared with those of our earlier study to investigate evidence of period changes.
Finally, we comment on longer-term trends in the light curves.

\section{Program Objects}

The program objects are listed in Table~\ref{object_list}, where we have identified them based upon their IRAS, Two Micron All Sky Survey (2MASS), and Gaia catalog numbers.  Also listed are their equatorial and galactic coordinates, apparent brightnesses, colors, and spectral types.  Several of them lie quite close to the galactic plane, as expected for intermediate-mass stars.
Their average {\it V} brightness ranges from 8 to 14 mag and most are very red, with ({\it B$-$V}) $\ge$ 1.8 mag. 
The spectral types range from F3 to G8, with a supergiant luminosity classification.  While these observed colors are much redder than expected for their spectral types, they are consistent with the presence of significant circumstellar and interstellar reddening.

\placetable{object_list} 

These 12 objects all display the properties of PPNe, including double-peaked SEDs, a compact nebula (all but IRAS 19500$-$1709 have been imaged with the Hubble Space Telescope, and about half, including IRAS 19500$-$1709, have resolved mid-infrared images). 
AFGL 2688 (the ``Egg Nebula'') is an exception, as it displays large bipolar lobes ($\sim$15$\arcsec$ end-to-end on the sky) 
illuminated by scattered light from the central star, and the star itself is obscured in visible light by a dark dust lane.
They have supergiant-like spectra, are somewhat metal-poor, are overabundant in the {\it s}-process elements, and are C-rich.  Their mid-infrared spectra show emission features characteristic of carbon dust.  More details of the properties of these 12 objects are given by \cite{hri10}.
The SEDs, Gaia distances, and derived luminosities for all of these except AFGL 2688 are given by \citet{kam22}. 
Photometric monitoring of a few of these has also been carried out by Arkhipova and collaborators \citep{ark09,ark10,ark11}.  
Some of these earlier data were included in one of our earlier studies \citep{hri13}.

\section{Observations and Data Sets}

\subsection{Our New VUO Observations}

We continued our observations of 11 of these 12 objects\footnote{IRAS 20000+3239 was not observed in the VUO-new observing program because of the increased exposure times used and limitations on available observing time.} from 2008 through 2018 at the VUO, using a new SBIG 6303 CCD camera.  
These we refer to as VUO-new observations.
This camera possessed advantages over our earlier camera by being larger in size and having an auto-guiding option.  The latter of these allowed us to take longer exposures than with our older detector, and thus we have improved the signal-to-noise ratio (S/N) for the fainter sources.  The former of these gave us options to change to better comparison stars, which we did in a few cases.
Observations were made through the Johnson {\it V} and Cousins {\it R}$_C$ filters, and, for the four brightest ones ({\it V} $<$ 10 mag), through the Johnson {\it B} filter also. 
However, due to a filter problem, there is a gap in {\it B} observations from 2015 June through 2016 September.

We carried out differential photometry of the sources, using three comparison stars 
for each object.    These are generally the same as in our earlier study of these objects, but not in all cases.   
The observations were reduced using  standard procedures in IRAF\footnote{IRAF is distributed by the National Optical Astronomical Observatory, operated by the Association for Universities for Research in Astronomy, Inc., under contract with the National Science Foundation.} cosmic ray removal, bias subtraction, and flat-fielding.  
Measurements of the brightness were made using aperture photometry, with an aperture of $\sim$5.5$\arcsec$ radius.
For AFGL 2688, we measured only the brighter northern lobe.    
The observations were transformed to the standard system through observations of Landolt standard stars \citep{land83,land92} and the application of linear color coefficients.

The primary comparison star (C1) in each PPNe field is listed in Table~\ref{std_comp}, along with its standard magnitude and color. 
These comparison stars were found to be constant at the level of $\pm$0.01 to $\pm$0.02 mag over the interval of observations.
In two cases (IRAS 05341+0852 and 07430+1115), we changed the primary comparison star (C1) from that used in our earlier study.    
To compare our previous differential data with the new data for those two stars, we empirically determined offsets from our measurements of the new and old comparison stars.  These are listed in the notes of Table~\ref{std_comp}.
We also use Table~\ref{std_comp} to record the number of observations of the program star in each filter.  These range from 41 to 395, with the winter objects having a smaller number of observations. 

\placetable{std_comp} 

In Table~\ref{std_diffmag} are listed the VUO-new standard differential magnitudes of each of the 11 PPNe for which we have new VUO observations, including the four that we had previously studied.  For IRAS 22223+4327 and 22272+5435, we had listed measurements from 2008$-$2012 in our previous paper \citep{hri13}, but list them here for completeness.  For IRAS 04296+3429 and 23304+6147, we used these observations in our recent study \citep{hri20a}, but the measurements were not listed there.
Uncertainties in the standardized differential measurements are approximately $\pm$0.005$-$0.010 mag except for IRAS 04296+3429, for which they are approximately $\pm$0.020 mag.

\placetable{std_diffmag} 

\begin{figure}\figurenum{1}\epsscale{1.20} 
\plotone{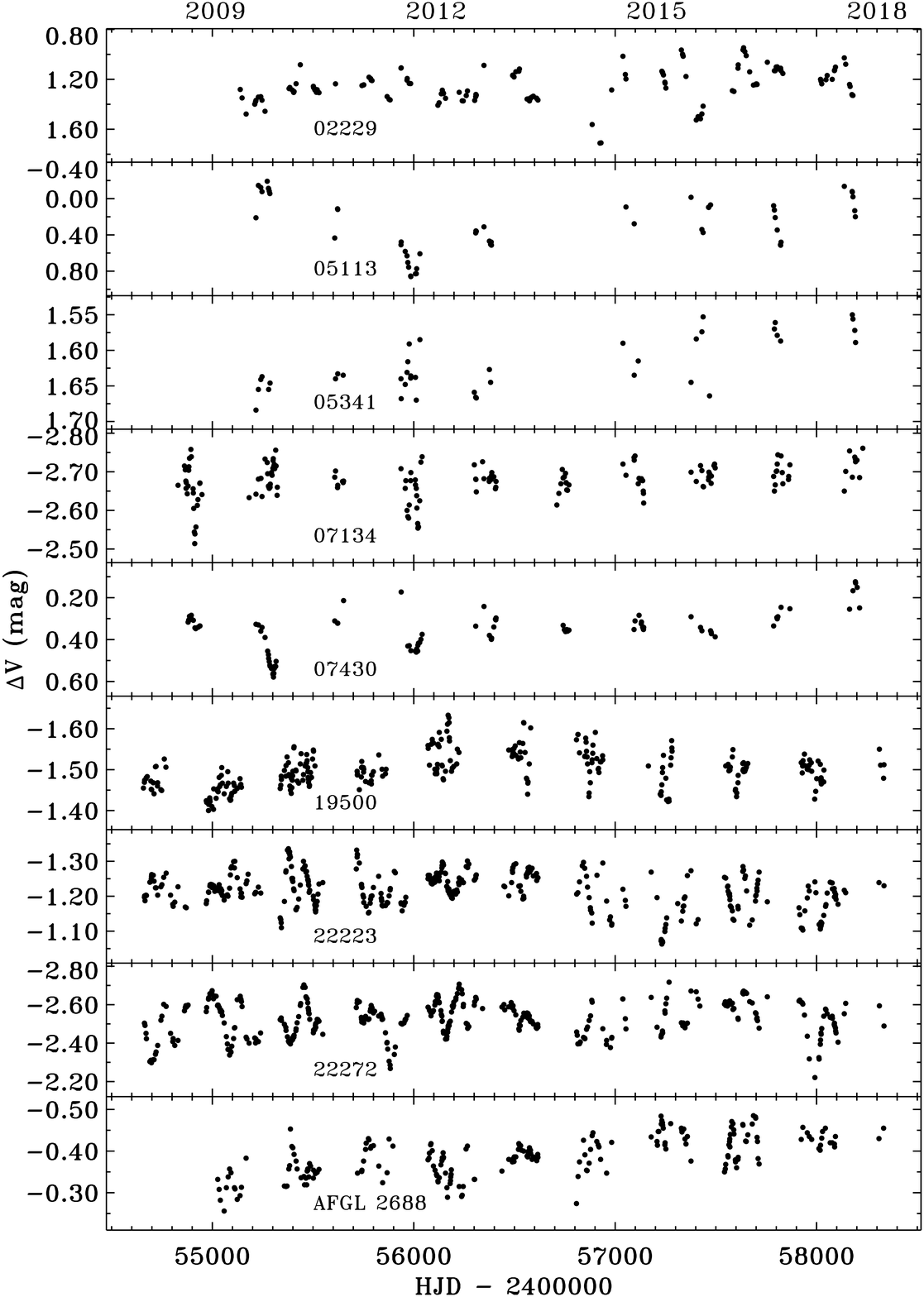}
\caption{The differential VUO {\it V} light curves from 2008$-$2018 for nine of the program objects.  
The uncertainties in the data are small, $\pm$0.005$-$0.010 mag.
Note that there are large differences among the brightness ranges of the objects (e.g., IRAS 05113+1347 and 05341+0852).
\label{fig1}}
\epsscale{1.0}
\end{figure}

In Figure~\ref{fig1} is shown the VUO-new light curves for nine of the objects, excluding IRAS 04296+3429 and 23304+6147, which we displayed in our recent study \citep{hri20a} and IRAS 20000+3239, for which no new VUO observations were made.  
One can see that all of the objects vary in brightness.  However the amount of change in the brightnesses varies greatly among the objects, ranging from a high of 1.02 in IRAS 05113+1347 to as low as 0.14 in IRAS 05341+0852.

\placefigure{fig1}

\begin{figure}\figurenum{2}\epsscale{0.80} 
\plotone{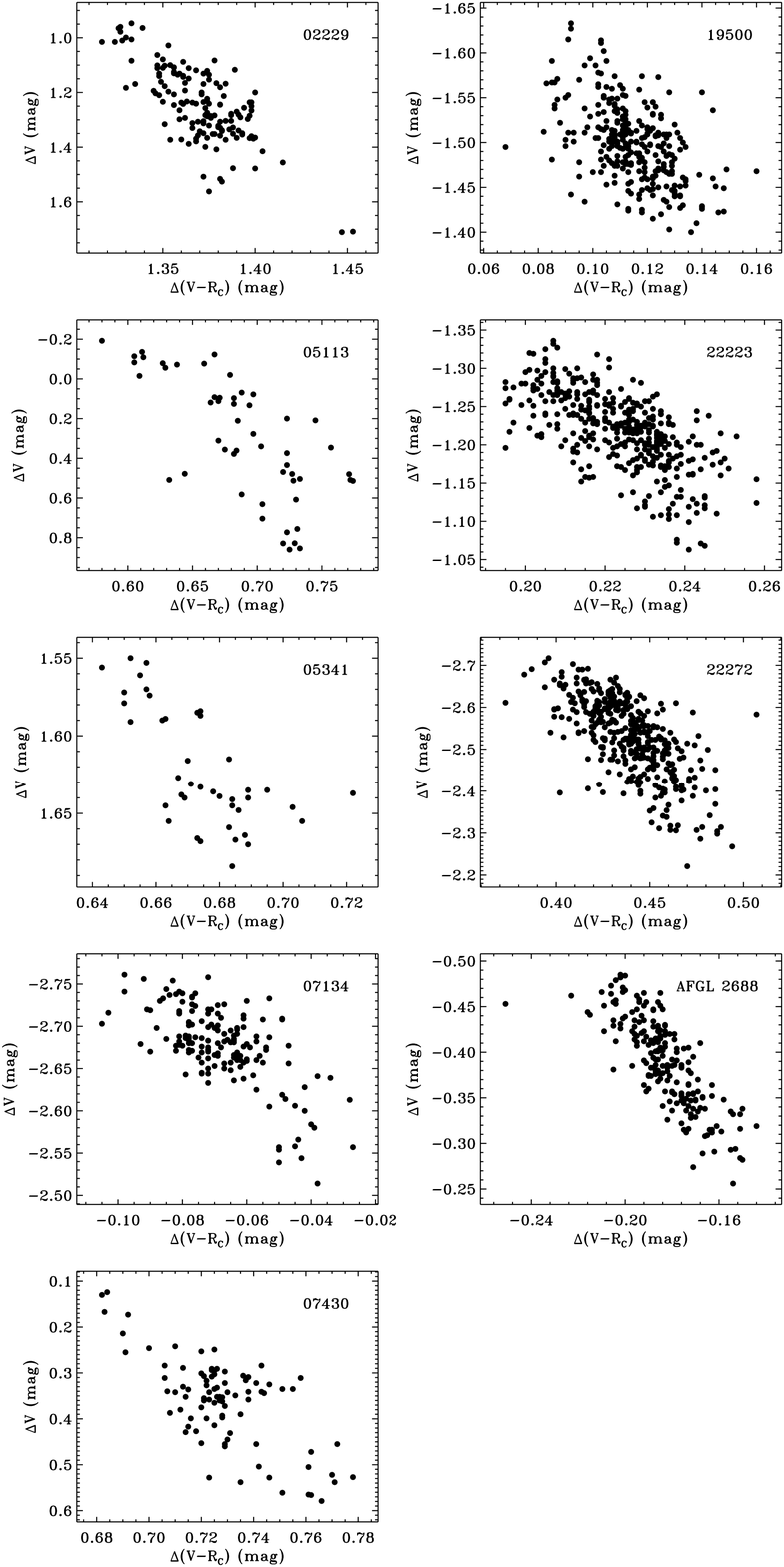}
\caption{The differential color curves for the target objects based on our new VUO data: $\Delta$({\it V$-$R$_C$}) versus $\Delta${\it V}.  There exists a clear trend of the stars appearing redder when fainter.
\label{color}}
\epsscale{1.0}
\end{figure}

The objects also vary in color, with a clear tendency to be redder when fainter.  This is shown in Figure~\ref{color}.
The range in ({\it V}$-${\it R}$_C$) color is as large as 0.17 mag for IRAS 05113+1347 to as small as 0.06 mag for several of the objects.

\placefigure{color}

\begin{figure}\figurenum{3}\epsscale{1.10} 
\plotone{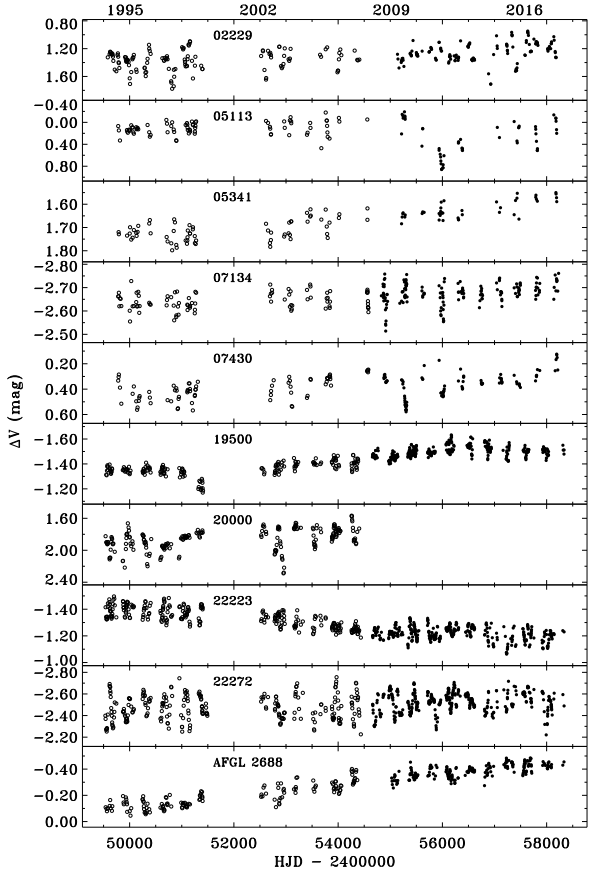}
\caption{The VUO-old (1994$-$2007) {\it V} light curves \citep[open circles;][]{hri10} plotted together with the VUO-new (2008$-$2018) {\it V} light curves from this study (filled circles).  In addition to seasonal variations, 
they allow us to see longer-term trends in the light curves.  
\label{fig4}}
\epsscale{1.0}
\end{figure}

We desire to investigate changes in the period and also the presence of long-term trends, especially periodic ones, that might indicate the presence of a binary companion.
To that end, we display in Figure~\ref{fig4} the combined VUO-new and VUO-old (1994$-$2007) {\it V} light-curve data of 10 of the objects.  
These light curves for IRAS 04296+3429 and 23304+6147 have been displayed by \citet[Fig. 2]{hri21}.
We note that in combining the VUO-old and VUO-new data from two detector-filter sets, there may be small offsets due to the neglect of second-order effects when standardizing these observations.  Most of these stars are very red ({\it B$-$V})$\ge$1.8), and so a small second-order coefficient can have effects at  the level of a few hundredths of a magnitude.  

\placefigure{fig4}

\subsection{ASAS-SN Data}
\label{asas-sn-data}

Photometric monitoring data of these objects are also available from the All-Sky Automated Survey for Supernovae \citep[ASAS-SN;][]{koch17}.  
This survey uses a series of cameras at five good astronomical sites, with observations made with the {\it V} and {\it g} (the Sloan Digital Sky Survey {\it g} filter, $\lambda$$_{\rm eff}$=477 nm) filters.  Typically, three consecutive, dithered observations were made, each of 90 s.  
The photometric measurements are made using aperture photometry, using a rather large radius of 16$\arcsec$. 
We combined these nightly individual measurements to determine an average magnitude, first eliminating data of inferior quality.  
In some cases for the {\it V} measurements and in all cases for the {\it g} measurements, observations were made with multiple cameras.  
For these cases, we detected small offsets between the measurements made with the different cameras and determined empirical offsets to correct these.
More details of these procedures are listed in a previous paper in this series \citep{hri21}.

ASAS-SN observations of a few of the objects began in the fall of 2013 and all of them began by 2015 April, beginning with the {\it V} filter.  These {\it V} measurements extend to 2018 November in all but two cases.
Similarly, {\it g} observations began for a few objects in the fall of 2017 and for all of them by 2018 September and continue to the present time.  We stopped our data gathering for this project with observations through 2021 April or May.  
For the three brightest objects, IRAS 07134+1003, 19500$-$1709, and 22272+5435, some of the {\it g} light-curve observations from the end of the observing interval (last $\sim$40 days) fell well below the level of the other measurements.  
Our suspicion is that this may be due to inadequate correction for saturation effects in the images, which begins at 10$-$11 mag.  
There is a saturation correction that is made in the reduction process, and, while it is helpful, it is not perfect \citep{koch17}.  For these bright objects ({\it V}=8$-$9 mag), with their low amplitudes of variation, this can be especially problematic.  We removed the obviously outlying data points by conservatively removing all measurements from the last $\sim$ 40 days of the above observing interval.  
Uncertainties in the ASAS-SN measurements of these objects are approximately $\pm$0.005$-$0.012 mag except for IRAS 04296+3429 and 20000+3239, for which they are approximately $\pm$0.020 mag.

\begin{figure}\figurenum{4}\epsscale{1.0} 
\plotone{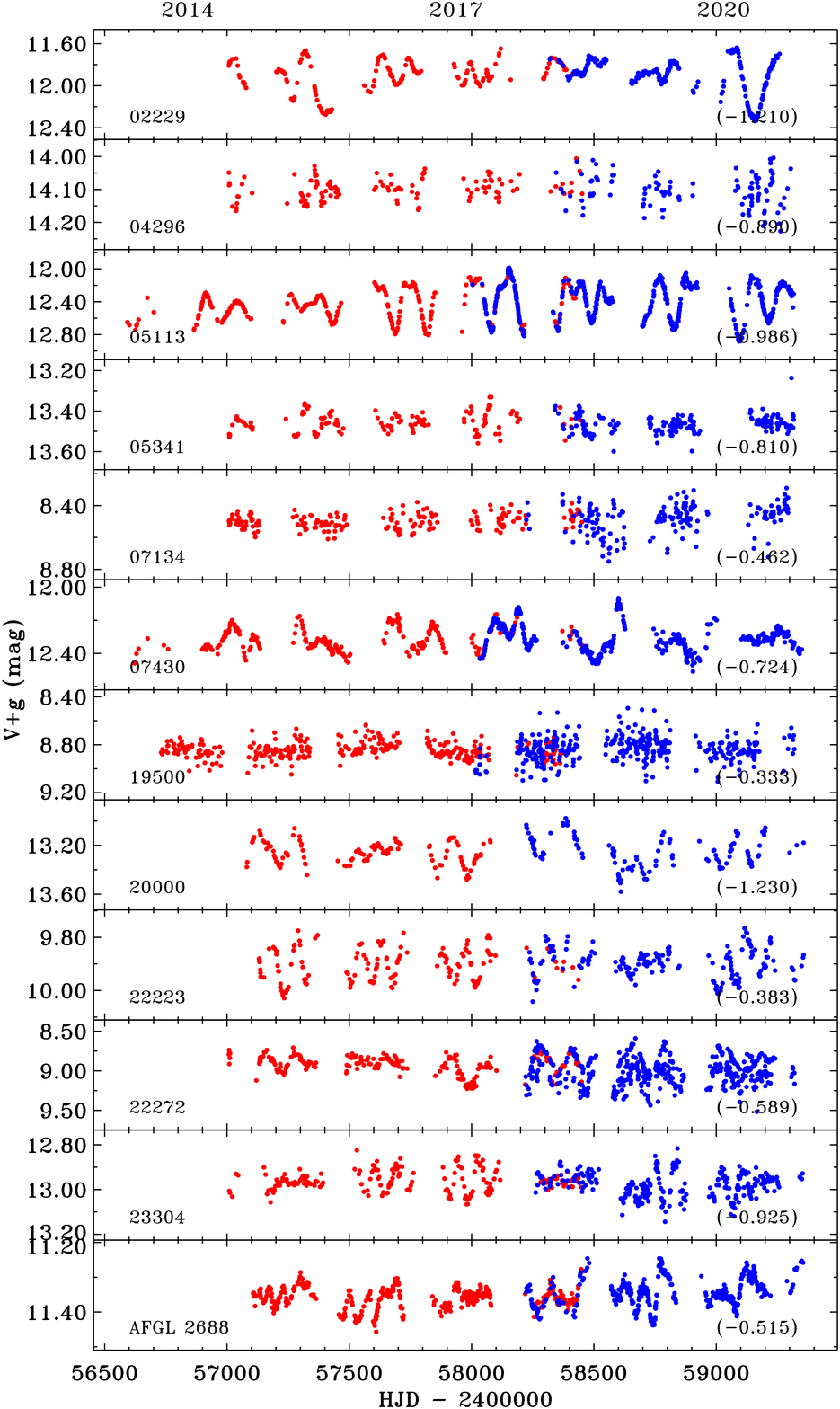}
\caption{The ASAS-SN {\it V} light curves (red) and {\it g} light curves (blue), with the latter offset to the level of the {\it V} light curve for the interval of overlap.  The amount of the offset in units of magnitudes is listed in parentheses. 
\label{fig3}}
\epsscale{1.0}
\end{figure}

This observing program resulted in all but one case with a temporal overlap of measurements with both filters during parts of one season (2018), and in parts of two seasons (2017 and 2018 ) in a few cases.  
For IRAS 20000+3239 there is no overlap.
Together, the ASAS-SN {\it V} and {\it g} data used in this study cover six or seven seasons of observations, typically four in {\it V} and three in {\it g}, and these are displayed in Figure~\ref{fig3}.
To make it easier to see the variability of each object in a single graph, we have offset the {\it g} measurements to agree with the {\it V} measurements by the use of an 
offset determined empirically from observations made during the temporal overlap.  
For almost all of the stars, one can visually see cyclical variability in the light curves.  

\placefigure{fig3}

We found in our analysis that, for the objects with longer periods ($>$80 days), the periods derived from the {\it g} light curves were often not in good agreement with those derived from the VUO-new and the ASAS-SN {\it V} light curves.  This we attribute to the shorter duration of the {\it g} light curves, compared to the other, longer light curves, and the presence of multiple periods in most of the light curves.  So, in practice, the {\it g} light curves did not contribute much to the period analysis for most of the objects.
However, for those with shorter periods, IRAS 07134+1005 and 19500$-$1709, the high-density {\it g} data, and the fact that the pulsation amplitudes are larger at shorter wavelengths, made these light curves particularly useful.

\section{Light Curves and Variability Study}

\subsection{Analysis Strategy}
\label{analysis}

With our new VUO-new data and the ASAS-SN measurements, we now have several additional light curves to analyze beyond what were available in our earlier study \citep{hri10}.
For each of the targets, we began with the analysis of the VUO-new {\it V}, {\it R}$_C$, and in four cases, {\it B} light curves, which cover 10 or 11 seasons, but typically with not as many data points in a season as the ASAS-SN data.
By contrast, the ASAS-SN {\it V} and {\it g} cover typically only four or three seasons, respectively, but they usually include many more data points (100$-$500) per filter, and certainly more per season.
For each object, we then compared the results of each of these analyses to see how consistent they were.  
The VUO-new data have the advantage of a longer time interval of observations, which can be particularly helpful in determining the period in cases where there are longer ({\it P}$\ge$100 days) or multiple periods.  The ASAS-SN data, with their shorter time intervals, are not as useful in those cases, but they can be particularly useful for objects with shorter periods ({\it P}$\le$40 days).
After assessing the results from the individual light curves, we combined the VUO-new and ASAS-SN {\it V} light curves for what we consider to be the most determinative analysis of the current value of the period. 
Since these two light curves overlap during four seasons, we were able to empirically determine an offset between our standardized differential VUO-new light curves and the ASAS-SN standard magnitude light curves.  The two data sets, when thus combined, show very good agreement in the resulting light curve.
For the fainter objects ({\it V} $\ge$ 10 mag), for which the VUO-new data have integration times of 30 to 45 min, we have given the VUO-new data points a higher weight (twice as much) in the analysis due to their better precision.
In the majority of cases, inspection of the VUO-new light curve showed changes in the median values from season to season.  To accommodate this, we removed trends and adjusted the data to the seasonal mean values before the period analysis, to focus on the pulsational periods.  We do note the cases where trends appear in the light curves and will discuss them separately.

These new period values were compared with the results from our earlier study of these objects, based on observations from 1994 to 2007 (VUO-old data), to see if any significant period changes have occurred.
Finally, we combined the VUO-old data with the VUO-new and ASAS-SN {\it V} data sets into a larger period analysis (1994$-$2018) to see what could be learned.

As in our previous studies, the main light-curve period-finding program that we used was PERIOD04 \citep{lenz05}.  
It is a commonly used program, easy to use, and accommodates multiple periods using a least-squares fitting routine.  
The most likely period is found using a Fourier analysis.  The light curve is then fit by a sine curve or curves, which works well with these relatively low-amplitude pulsators (typical variations of 0.2-0.4 mag in a season), and PERIOD04 has the advantage that it can fit multiple periods simultaneously.  
A period is judged to be significant if it meets the criterion of a signal-to-noise ratio (S/N) $\ge$ 4.0 \citep{bre93}. 
To confirm these results for the dominant period of each, we analyzed the data with several other period-search programs \citep[found in Peranso;][]{paun16} and found similar results. 
The uncertainties quoted are those that are derived from the formal PERIOD04 least-squares analysis, and are typically $\pm$0.1$-$0.3 days.  
A more conservative estimate might be about five times larger.

\begin{figure}\figurenum{5}\epsscale{1.2} 
\plotone{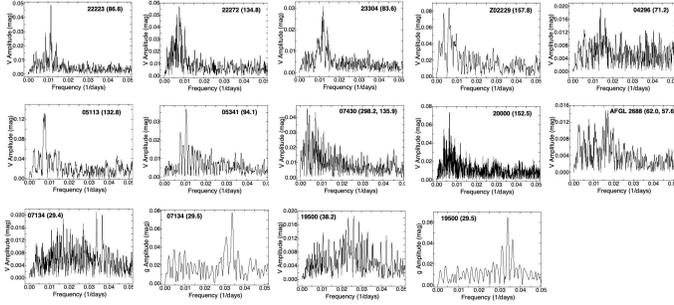}
\caption{ The frequency spectrum of the normalized {\it V} light curve of each of the objects.
The period, in days, is listed in parentheses. 
For the two shorter-period objects, IRAS 07134+1005 and 19500$-$1709, we have also the frequency spectra of the larger-amplitude ASAS-SN {\it g} light curves.  These are shown in the bottom row.
\label{FreqSpec}}
\epsscale{1.0}
\end{figure}

Below are presented the analyses of the light curves for each of the program objects.
We begin with a brief description of the light curves and then discuss the results of our analyses of them.
The frequency spectrum for each object, based upon the analysis of its light curve, is displayed in Figure~\ref{FreqSpec}.
The results of the analyses of the combined, adjusted {\it V} light curves are given in Table~\ref{periods}, which lists the filter, years of observations, data sets, number of observations, and the period, amplitude, and phase for each significant period.  Also listed are the standard deviations of the observations from the sine-curve fit.
The light curves fitted by these results are displayed in Figure~\ref{LC_fits}.   We will evaluate the fits as we discuss the individual stars.

\placefigure{FreqSpec}

\placetable{periods} 

\placefigure{LC_fits}

\subsection{Complications in the Light Curves}
\label{compl}

However, before we begin to investigate the individual objects, we briefly discuss some of the known complications in the light curves of PPNe, with an eye to the impact that they have on the analysis of their light curves.  This is based on our 25 yr of observations and analyses.
The light curves often show very large difference in the range of variations from season to season.  
There often exist two or even three significant periods in the light curves of these stars.  
Even when fitted with multiple periods, the resulting parameters often do not fit the light-curve amplitudes well, suggesting that the amplitudes are not constant.
This is not surprising, since it has been shown by analyses of high-resolution spectra that shocks \citep{leb96,zacs16} and possibly large-scale convective motions \citep{zacs21} exist in some of these objects.
These are expected to impact the light curves.
These complications need to be kept in mind when assessing the goodness of the light-curve fits.
These things are especially problematic in the light curves of stars of shorter periods, for in them the amplitudes of the pulsation are smaller.
In contrast, for the longer-period pulsators, the light-curve changes occur more slowly and the amplitudes are larger, with the consequence that the complications are of relatively smaller significance.

\subsection{Individual Objects}

{\it IRAS 22223$+$4327} $-$ 
In our earlier studies, we documented the cyclical behavior of the light curve of IRAS 22223+4327 and its changing amplitude, which suggested multiple periods \citep{hri10,hri13}.
Also seen were changes in the mean brightness levels in each season.  
Cyclical variations can be seen visually in the VUO-new and ASAS-SN data, with varying amplitudes that reach a maximum {\it V} range of 0.26 mag in 2015$-$2016.

Good agreement is found 
among the periods determined in the individual adjusted light-curve analyses, with a strong dominant period of 87 days and a secondary period of 91 days.
Analyzing the adjusted VUO-new and ASAS-SN combined {\it V} light curve resulted in a dominant period of {\it P}$_1$ = 86.6$\pm$0.1, a secondary period of {\it P}$_2$ = 91.2$\pm$0.1 days, and a tertiary period of {\it P}$_3$ = 116.1$\pm$0.2 days.
The results of this combined, adjusted {\it V} light curve are given in Table~\ref{periods}, which lists the filter, years of observations, data sets, number of observations, and the period, amplitude, and phase for each significant period.  Also listed are the standard deviations of the observations from the sine-curve fit.
The fit is good, with a standard deviation of 0.028 mag, and is displayed in Figure~\ref{LC_fits}.  

An analysis of the adjusted VUO-old {\it V} light curve yields periods of {\it P}$_1$ = 87.7$\pm$0.1 and  {\it P}$_2$ = 90.8$\pm$0.1 days, similar to what we find in the newer data.
Inspection of the combined VUO-old and VUO-new 
observed {\it V} light curve (Fig.~\ref{fig4}) shows a monotonic decline of 0.18 mag from 1994 to 2008-09, with perhaps a bit of an inflection there and then a small decline beginning in 2015.  Analysis of the entire combined {\it V} light curve from 1994 to 2018, including ASAS-SN data and adjusted to seasonal mean values, reveals a dominant period of 87.1$\pm$0.1 days.

{\it IRAS 22272$+$5435} $-$ 
As in our earlier studies of IRAS 22272+5435, these VUO-new observations display a clear cyclical behavior with changing amplitudes and variations in the seasonal brightness levels.
The VUO-new amplitudes reach a {\it V} maximum in 2011-2012 of 0.35 mag. 
Similar variations are seen in the ASAS-SN data, reaching a maximum {\it V} amplitude in 2017-2018 of 0.46 mag, but there is more scatter, especially in the {\it g} light curve.  This may be due to saturation effects.  
As we have previously noted, the last ~40 days of observations in the {\it g} filter were deleted because they were significantly lower than other observations.

Analysis of the VUO-new data, when adjusted to the seasonal mean values, revealed a consistent dominant period of 132 days and a secondary period of 121 days in the {\it V}, {\it B},  and {\it R}$_C$ data, with an additional period of 113 days in the {\it V} data.  These give good fits to the light curves.  
The adjusted ASAS-SN data, covering only four {\it V} or three {\it g} seasons, gave a dominant period of 124 days and a secondary period of 113 days {\it V}. 
These are relatively close to the VUO-new results, and rather than infer a recent period change, we attribute the difference to the smaller time interval of the data and perhaps the lower reliability due to the saturation concerns in the ASAS-SN data for this bright ({\it V}=9.0 mag) star. 
However, this should be followed up by continued monitoring of this object to see if there is indeed a confirmed decrease in the period of variation.
We then did an analysis of the combined VUO-new and ASAS-SN {\it V} light curve, which resulted in periods of 134.8, 121.6, and an additional period of 154.0 days.
These periods and their resulting amplitudes give a reasonably good fit to the times of maximum or minimum, but in some seasons, the fits to the changing observed amplitudes are not good, especially in the last season.
The dominant period from this combined VUO-new and ASAS-SN light curve is slightly longer, by $\sim$2 days (1.5\%) than that found in the individual light curves over this time interval.
Analysis of the observed data, not adjusted to seasonal means, resulted in an additional significant period of 1.8 yr.  When we include the VUO data, the longer period value of 3.2 yr emerges instead.  Thus there is a suggestion of an additional period on the order of 2$-$3 yr in the data.  This will be discussed later (Sec.~\ref{trends}).

The VUO-old {\it V} data yielded periods of 132 and 125 days, so the dominant period was the same as that found in the newer individual light curves.
The analysis of the combined VUO-old, VUO-new, and ASAS-SN {\it V} light curve, adjusted to the median values, resulted in a dominant period {\it P}$_1$ = 131.9$\pm$0.1 days, with {\it P}$_2$ =  127.2$\pm$0.1 days, and an additional period of 154.3$\pm$0.2 days.

{\it IRAS 23304$+$6147} $-$ 
In a recent study, we analyzed the VUO-new data, which resulted in  {\it P}$_1$ = 83.8 $\pm$ 0.1 and  {\it P}$_2$ = 70.2 $\pm$ 0.1 days \citep{hri20a}. 
The ASAS-SN {\it V} light curve shows clear cyclical variability in seasons two and three, but very little variability in seasons one and four.  
Analysis of the data from seasons two and three results in a period of 84.0 $\pm$ 0.3 days.
For consistency with this study, we combined the VUO-new and the ASAS-SN {\it V} light-curve data, adjusting them to seasonal median values.  
The analysis of this data set resulted in a dominant period of 83.6$\pm$0.1 days and a secondary period of 87.3$\pm$0.1 days, which serves basically to modulate the amplitudes.

Analysis of the earlier VUO-old {\it V} data yielded a period of 84.6 $\pm$ 0.1 days, similar to the more recent value.
Combining all three of the {\it V} light curves, adjusting them to season-median values and analyzing them, resulted in a dominant period of 84.4 days and a secondary period of 80.2 days, again with the secondary basically modulating the amplitudes.  Analyzing only the VUO-old and VUO-new data without the recent ASAS-SN data resulted in a dominant period of 84.5 days but a secondary period of 70.4 days.
Thus we derive from the newer data a period of 83.6$\pm$0.1 days, with a secondary period that appears to serve primarily as an attempt to fit the changing amplitudes of the light curves.  This is close to the period found from the VUO-old data.

{\it IRAS Z02229$+$6208} $-$ 
The VUO-new data show large variations in some seasons, reaching 0.70 mag ({\it V}), but smaller ones in others, $\sim$0.25 mag.  The variations in {\it R}$_C$ are similar but slightly smaller.
The overall variation is 0.77 mag in {\it V}, and there appears to be an overall increase in brightness during this observing interval, particularly beginning in the 2014$-$2015 season.
There are additional variations in the overall season-to-season brightnesses.
A similar pattern of variations was seen in the earlier VUO light curves.  
The ASAS-SN {\it V} light curve covers parts of five seasons, extending one season later than the VUO {\it V} light curve, and shows similar variations.
These two {\it V} light curves show good agreement in the seasons of overlap.
The ASAS-SN {\it g} light curve covers most of three seasons with low-amplitude variations in two of the seasons and a very large cyclical variation in the third season of 0.7 mag. 

Analyses of the individual VUO-new {\it V}, {\it R}$_C$, and ASAS-SN {\it V} light curves yielded similar values of $\sim$157 days.
This is superimposed on a general increase in brightness. 
The combined VUO-new and ASAS-SN {\it V} light curve was analyzed by first removing the general trend and then adjusting the residuals from the trend to seasonal median values.  This resulted in similar period of {\it P}$_1$ = 157.8$\pm$0.4 days, with evidence of a second period of 261.2$\pm$1.1 days, which improves the fit slightly.  
The fit to this solution is shown in Figure~\ref{LC_fits}. 
The amplitude and phasing are off a bit at times, but it is a reasonable representation except for the very last season of ASAS-SN data. 

The period found in the 1994$-$2007 light curves is 153 days, slightly shorter than that of the newer light curves.
Combining the VUO-old with the VUO-new and ASAS-SN {\it V} light-curve data results in {\it P}$_1$ = 154.6$\pm$0.1 days, along with a second period of 284.0$\pm$0.4 days.

{\it IRAS 04296$+$3429} $-$ 
This object was also recently analyzed by \citet{hri20a}.   They found that the VUO-new light curve continued the general trend of increasing brightness seen in the VUO-old data, increasing by 0.18 mag ({\it V}) from 1994 to 2017.  The data were adjusted to correct for this.
The period of the VUO-new {\it V} light-curve data is 71.3 $\pm$0.1 days. 
The ASAS-SN {\it V} light curve possesses periods of 67.1 and 76.2 days.
Combining the VUO-new and ASAS-SN {\it V} data results in {\it P}$_1$ = 71.2$\pm$0.1 days and two additional significant periods of 68.8 and 76.7 days.

The VUO-old data have an approximate period of 71 days, in agreement with the dominant period found in the recent data.  Analyzing all three {\it V} light curves results in {\it P}$_1$ = 71.1$\pm$0.1 and {\it P}$_2$ = 68.9$\pm$0.1 days.

{\it IRAS 05113$+$1347} $-$ 
The VUO-new data show evidence of cyclical behavior, but are few in number (53 in {\it V} and 62 in {\it R}$_C$) and do not go through a complete cycle in any one season.  
The observed  range in brightness in a season is relatively large, up to 0.44 mag ({\it V}).
The overall {\it V} light curve shows a decline in brightness from 2010$-$2012 of $\sim$0.6 mag and then a slower rise up to the original brightness and covers a range of 1.05 mag peak-to-peak.  The {\it R}$_C$ light curve is similar in appearance.
The ASAS-SN data, by contrast, are very numerous and reach even larger amplitudes (0.72 mag; {\it V}) in season.  Visual inspection reveals a cyclical variation of 120$-$130 days in a season, with varying amplitudes among the seasons.

Analyzing the individual ASAS-SN {\it V} (four full seasons and parts of two others), {\it g} (most of four seasons), and even the combined {\it V} and {\it g} (using an empirically  determined offset) light curves yielded relatively similar periods of $\sim$125 and $\sim$136 days.  
The VUO-new {\it V} and {\it R}$_C$ light curves, with many fewer data points, yielded a period of 96 days.
We then combined the VUO-new and the ASAS-SN {\it V} data, adjusted to seasonal median values to compensate for the sharp decline mentioned above and removing the seasons with fewer than six data points.
This was analyzed and yielded 
 {\it P}$_1$ = 132.8$\pm$0.1, {\it P}$_2$ = 121.4$\pm$0.1, and {\it P}$_3$ = 135.5$\pm$0.1 days.  
 The third period appears to serve mainly to modulate the amplitudes of the sine-curve fit.  
The fit is reasonably good, given the large variation in seasonal amplitudes.
 
The VUO-old {\it V} light curve showed similar portions of cyclical behavior as does the VUO-new, although with more data points (78) and a larger peak-to-peak range of 0.67 mag in a season.  The VUO-old data, however, did not show the seasonal decline followed by the slow increase in brightness seen in the newer data. 
Analysis of the VUO-old data yielded  {\it P}$_1$ = 133.0$\pm$0.3 days ({\it V}) and 137.4$\pm$0.4 days ({\it R}$_c$).
Lastly, we combined all of the {\it V} data, VUO-old, VUO-new, and ASAS-SN, adjusted to the seasonal median values, ignored a few seasons with few data points, and analyzed this complete light curve.  
This resulted in two strong periods of  {\it P}$_1$ =  121.6$\pm$0.1 and  {\it P}$_2$ =  137.4$\pm$0.1 days, and an additional period of  {\it P}$_3$ =  131.1$\pm$0.1 days. 
Thus, there appear to be two strong periods of 133 and 121 days in the newer 2008$-$2018 {\it V} data, while the older 1994-2007 {\it V} data shows a single period similar to the first of these.
 
The decline seen in the 2011$-$2012 data bears a resemblance to the similar decline seen in the VUO-old light curve of IRAS 19500$-$1709 in 2009 (see Fig.~\ref{fig4}).  This was ascribed to variable opacity in the circumstellar envelope of that star, perhaps due to a recent mass ejection event or the motion of a dust cloud across the line of sight \citep{hri10}.
However, the ({\it V$-$R}$_C$) color of IRAS 05113+1347 does not get systematically redder during this dimming episode.

{\it IRAS 05341$+$0852} $-$ 
The VUO-new light curves had sparse observations (41 in {\it V} and 47 in {\it R}$_C$).  They continued the trend of increasing brightness seen in the earlier VUO data, with an apparent increase of 0.08 mag in {\it V} over this new observing window. 
The light variations are not large, with a maximum of 0.11 mag in a season and 0.13 mag over the entire observing interval in {\it V}.
The ASAS-SN {\it V} light curve covers portions of five seasons with many more observations (138) and shows some larger variations with in a season, reaching a maximum of 0.23 mag.  Visual inspection reveals cyclical behavior, with a period of 90$-$100 days.  
The ASAS-SN {\it g} light curve covers three seasons.  Cyclical behavior is seen only in the first season, 2018$-$2019, the one season of overlap with the ASAS-SN {\it V} light curve. 

No significant periodicity is found in either of the VUO-new {\it V} or {\it R}$_C$ light curves.
Analysis of the ASAS-SN {\it V} light curve results in a period of 93.5 days over the five years of observations.  The {\it g} light curve is not periodic over the three years of observations, but an analysis of the first year alone results in a period of 90 days.
An analysis of the combined VUO-new and ASAS-SN {\it V} light curve was carried out, first removing the increasing trend of brightness but not adjusting to the seasonal median values due to the small number of observations in many of the VUO-new observing seasons.
This resulted in a period of 94.0$\pm$0.1 days.  
This gives a reasonably good fit, as shown in Figure~\ref{LC_fits}.

The VUO-old {\it V} light curves possesses a period of 93.9 days, and combining all three {\it V} light curves results in a period of 93.8 days.  Thus, the period appears to be unchanged over this observing interval.

{\it IRAS 07134$+$1005} $-$ 
The VUO-new observations show some evidence of cyclical variability in a season with a relatively shorter cycle length of 40$-$60 days, but with lots of scatter.  
The range of variability varies, reaching a maximum of 0.24 ({\it V}) and 0.32 mag ({\it B}) in the first season.  Those seasons with the larger range show the most clear evidence of cyclical variability.
Similar variations are seen in the ASAS-SN {\it V} light curve, which covers four and a half seasons.  .
The ASAS-SN {\it g} light curve shows much larger variations, $\sim$0.4 mag in a season, and a very clear cyclical variability of about 30 days.  

The analysis was carried out with the seasonally adjusted light curves.  No significant period was found in the VUO-new {\it BVR}$_C$ light curves, although they all indicated almost significant periods of 52.9 days.  
The period of 29.5 days was found in both of the ASAS-SN light curves, and was particularly strong in the larger-amplitude {\it g} light curve.  
We proceeded to analyze the combined VUO-new and ASAS-SN {\it V} light curve, and found {\it P}$_1$ = 29.4$\pm$0.1, {\it P}$_2$ = 36.5$\pm$0.1, and perhaps a third period of 58.0 days. 
The VUO-old data possess periods of 38.9 and 44.4 days. 
We also examined the VUO-old data with {\it P} = 29.4 days, but this shorter period did not fit the older data.

\begin{figure}\figurenum{7}\epsscale{1.2} 
\plotone{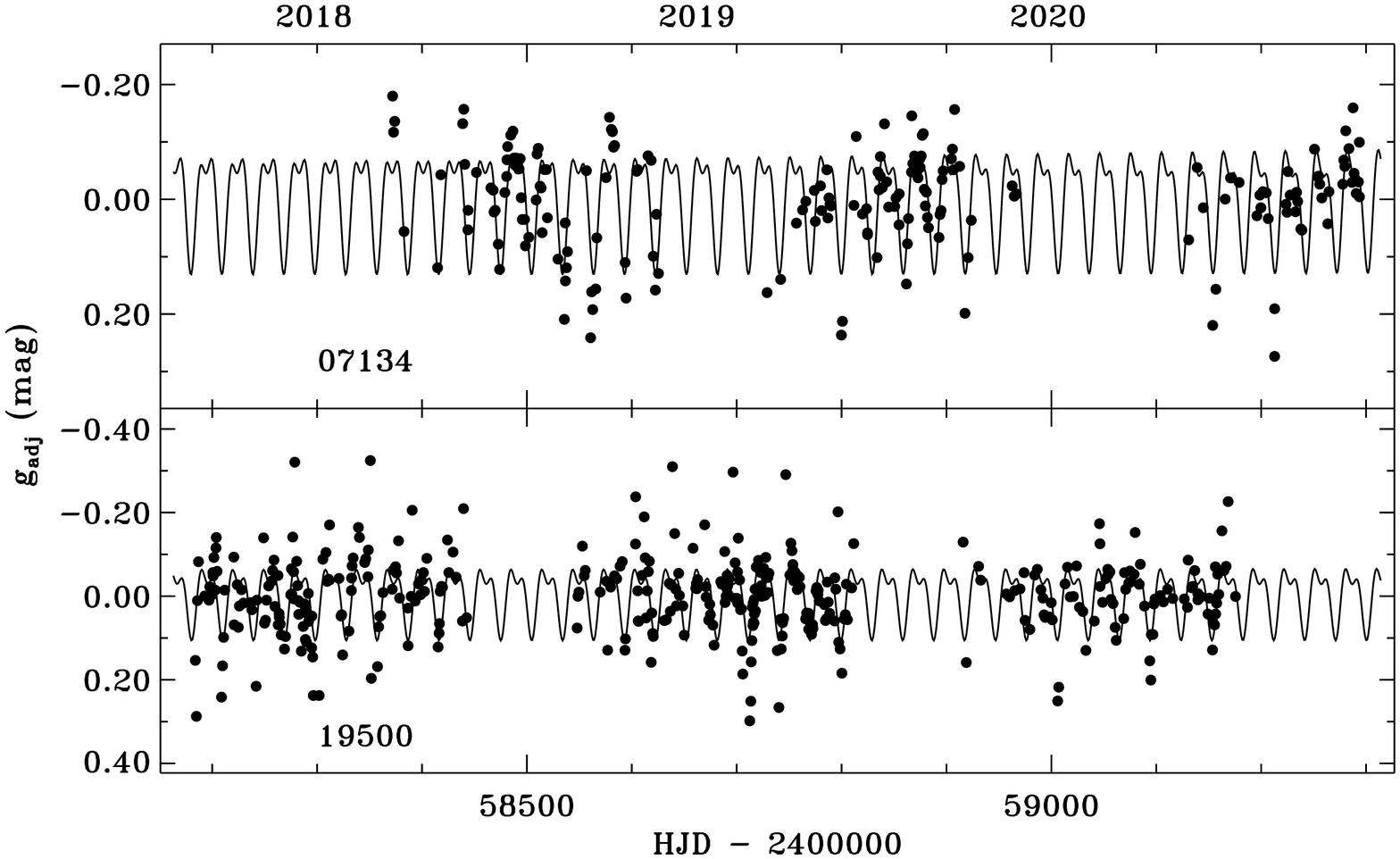}
\caption{The ASAS-SN {\it g} light curves of the shorter-period objects ({\it P}$\sim$30$-$40 days), normalized as described in the text, and
fitted by the periods, amplitudes, and phases recorded in Table~\ref{periods}.   
The data uncertainties are small, $<$0.005 mag.
\label{LC_fits-2}}
\epsscale{1.0}
\end{figure}

\begin{figure}\figurenum{6}\epsscale{1.2} 
\plotone{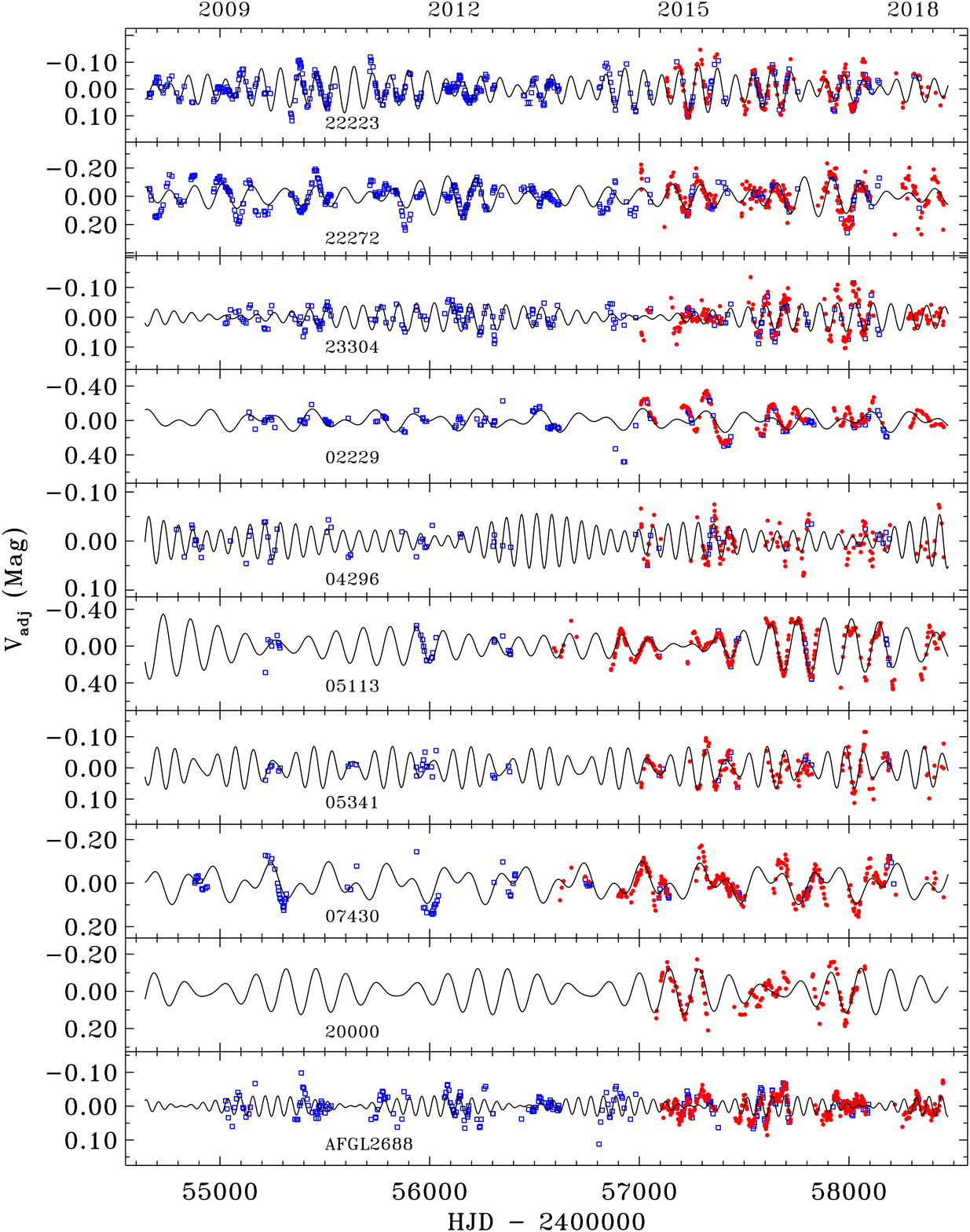}
\caption{The combined {\it V} light curves of the longer-period objects ({\it P}$\sim$60$-$160 days), based on VUO-new (blue squares) and ASAS-SN (red circles) data, and normalized as described in the text.
These are fitted by the periods, amplitudes, and phases recorded in Table~\ref{periods}.   
The data uncertainties are small, $\pm$0.005$-$0.010 mag, except for IRAS 04296$+$3429 ($\pm$0.015 mag).
\label{LC_fits}}
\epsscale{1.0}
\end{figure}

In this case, as in that of the other short-period object, IRAS 19500$-$1709, the fit of the pulsation analysis to the observations does not appear to be very good.
This is in part due to the small amplitude of the pulsation, 0.02 mag, and the presence of physical shocks \citep{leb96}.   
We have chosen to display instead the fit to the larger-amplitude (0.08 mag) ASAS-SN {\it g} light curve.  
In addition to a dominant period of 29.5$\pm$0.1 days, there is a second significant period of 14.8$\pm$0.1 days, half that of the main period.  
This appears to serve the purpose of modulating the shape and amplitude of the light curve. 
The fit for this {\it g} light curve over three seasons is shown in Figure~\ref{LC_fits-2}, and one can see the agreement amidst the scatter in the data.  
We have also included the frequency spectrum for this {\it g} light curve in Figure~\ref{FreqSpec}.

\placefigure{LC_fits-2}

This is one case in which there is strong evidence of a period change.  It appears that the period decreased from about 40 days in the interval 1994$-$2007 to 29 days in the more recent 2015$-$2021 time interval.

{\it IRAS 07430$+$1115} $-$ 
These VUO-new observations show patterns of variations similar to those seen in the VUO-old observations, with clear indications of cyclical variations but not enough observations in a season to visually estimate a period.  
In contrast, visual inspection of the ASAS-SN data, with many more observations in a season, clearly shows cyclical patterns.  However, the timings between adjacent maxima and adjacent minima vary over a range of 80$-$180 days, indicating multiple periods.  Along with this, there appear to be variations in the mean light level between the different seasons.
The maximum variations in a season reach 0.28 mag ({\it V}). 
The combined VUO-new and ASAS-SN {\it V} light curve from 2009$-$2018 varies over a total range of 0.45 mag, with a general increase in brightness of $\sim$0.2 mag.

We analyzed the individual light curves, adjusted to their seasonal mean values to remove the seasonal variations and the general increase in brightness. 
A period of 136 days is seen in the adjusted VUO-new {\it V} and {\it R}$_C$ light curves and in the ASAS-SN {\it V} light curve.  
When combined, the adjusted VUO-new and ASAS-SN {\it V} light curve yielded a primary period of 135.9$\pm$0.2 days, which fit the timing of the extrema of the light curve, and an additional period about twice that value, 298.2$\pm$0.8 days, which helped to fit the changing amplitudes of the pulsation.  
This solution gives a reasonable fit and is shown in Figure~\ref{LC_fits}.

The VUO-old {\it V} data, when adjusted, yielded periods of 136 and 149 days, the former being the same period as found in the newer light curves.
Similarly, analysis of the adjusted, combined VUO-old, VUO-new, and ASAS-SN {\it V} light curve, resulted in a dominant period {\it P}$_1$ = 136 days, with {\it P}$_2$ =  293 days.  
Thus, the dominant period of 136 days appears secure and appears not to have changed during this this overall observing interval of 1994$-$2018.

{\it IRAS 19500$-$1709} $-$ 
The VUO-old observations for this object showed a sudden drop in brightness in 1999, then slowly recovered in brightness over the next four seasons.  The data from 2008$-$2018 show that the brightness continued to increase by 0.11 mag ({\it V}) from 2008 to 2012, peaked, and then decreased in brightness by 0.07 mag ({\it V}) by 2018. 
Visual inspection of the VUO-new data from 2008$-$2018 shows indication of a cyclical behavior with period of $\sim$35$-$45 days, which is seen particularly in the seasons with the largest range in variation.
The ASAS-SN data show a much larger range of variation, at least part of which we attribute to scatter due to the saturation affects ({\it V}=8.7 mag).

In light of the seasonal changes in the brightness level, we first adjusted the VUO-new data to the median value of each season before we analyzed them.  
We omitted seasons with small numbers of observations. 
The periodogram analysis VUO-new {\it V} and {\it R}$_C$ data yielded a period of 38.2$-$38.3$\pm$0.1 days, with an average residual of 0.030 mag ({\it V}).  
When we compared the VUO-new and the ASAS-SN {\it V} light-curve data, the agreement was only fair, with the ASAS-SN data showing a lot more scatter.
No significant period was found in the ASAS-SN {\it V} light curve.  
With these two facts in mind, we decided not to combine the VUO-new and ASAS-SN data for analysis, but use only the results from the VUO-new data.

The VUO-old data were reanalyzed, and values of 36.8 and 44.1 days were found in the {\it V} data.  No significant period was found in the VUO-old {\it R}$_C$ data.
Analyzing the combined VUO-old and VUO-new data resulted in a well-determined period of 38.3 days in each of the {\it V} and {\it R}$_C$ data sets.  Searching for a second period in the data revealed a value of 45.0 days in the combined {\it R}$_C$ data, similar to that found in the older {\it V} data, so perhaps that might be real. 
Thus, the VUO observations indicate a period of 38.3 days, with little indication of change from the earlier and present observations. 
As with the other short-period, low-amplitude object IRAS 07134+1005, the fit to the {\it V} light curve does not look very good, although the period is significant, and thus we have not displayed it.
However, the ASAS-SN {\it g} data from 2018-2020 indicate a much shorter period of 29.5$\pm$0.1 days.
The frequency spectrum for this is included in Figure~\ref{FreqSpec}.
This light curve has a relatively large amplitude (0.07 mag).
It also has a weaker secondary period of half that value, and these together give a reasonably good fit to the observations, albeit with a large average residual (0.067 mag).  
This is shown in Figure~\ref{LC_fits-2}.
We suspect that the large residuals in the ASAS-SN data are due in part to the limited accuracy in the corrections for the saturation of this object, in addition to the real complications in the light curves.
We examined the VUO-new {\it V} light curve with this period of 29.5 days, but it was not a good fit to this data.
Continued observations are needed to investigate if this suggestion of a recent change from 38 days to a shorter period of 29 days is real.

{\it IRAS 20000$+$3239} $-$ 
The earlier light curve clearly showed cyclical variability of reasonably large amplitude (up to 0.6 mag in {\it V}) on some seasons and only small variations on other seasons \citep{hri10}.  As noted, this object was not included in the VUO-new observing program due to limitations in the observing time available.
The ASAS-SN light curves also show cyclical variability with changing amplitudes.  They cover three seasons in each of the {\it V} and {\it g} filters with no temporal overlap.  Season 2 in the {\it V} filter is significantly different from the others, with smaller variations on a shorter time scale superimposed on a general increase in brightness over that season.

Analysis of the ASAS-SN {\it V} light curve resulted in periods of 154 and 133 days with all of the data and 157 days if we omitted season 2.  The {\it g} light curve resulted in a period of 138 days.  Since these cover only three seasons each, their reliability is not high given the changing nature of the light curves. 
With that in mind, we next combined the ASAS-SN and the VUO-old {\it V} light curves, the latter of which contained 213 observations made over a range of 14 seasons.    
Analyzing this adjusted light curve resulted in {\it P}$_1$ = 152.5$\pm$0.1 and {\it P}$_2$ = 130.4$\pm0.1$days. This gives a reasonably good fit to the combined light curve.
This is in good agreement with the VUO-old results of 152.7$\pm$0.3 days.  
Thus, there is no evidence to suggest a period change.

{\it AFGL 2688} $-$ 
The VUO-new observations from 2009$-$2018 show the trend of increasing brightness seen on the earlier VUO-old observations.  
In the new data, we see an increase of $\sim$0.10 mag ({\it V}) from 2009 to 2015, and then an apparent leveling off in brightness through 2017.  
The data within a season show a cyclical variation, but with varying cycle lengths.
The ASAS-SN {\it V} data cover four seasons, with clear cyclical variations in the first two, in good agreement with the contemporaneous VUO-new data.

For the analyses, we adjusted the data to the seasonal median values to remove the trend in the data and other seasonal variations. 
The VUO-new data gave similar periods of 62 and 58 days in both the {\it V} and {\it R}$_C$ light curves.
The ASAS-SN {\it V} light curve had some longer periods to account for trends within the seasons and also included a period of 58 days.
The analysis of the combined VUO-new and ASAS-SN data resulted in periods of 62.0$\pm$0.1 and 57.6$\pm$0.1 days.  
This gives a reasonable fit to the light curve and is shown in Figure~\ref{LC_fits}. 

We note that these values differ greatly from the uncertain value of 93 days determined from the VUO-old {\it V} light curve from 1994$-$2007.  While these cover a longer interval of time, there are far fewer data points than in the VUO-new or the ASAS-SN light curves.  We also note that the VUO-old {\it R}$_C$ light curve yielded a period of 62 days.
Rather than concluding that the light-curve period has changed from 93 to 62 days, we think it more likely that the earlier {\it V} light-curve period is not reliable.  An analysis of the combined VUO-old, VUO-new, and ASAS-SN {\it V} light curve results in two similar-strength periods of 62 and 58 days.

\section{Discussion}

All 12 of the objects included in this study are found to vary periodically in brightness, with almost all of them possessing a second significant period.  The primary periods range from 29$-$158 days.  In most cases, the secondary periods are close to the primary ones, with {\it P}$_2$/{\it P}$_1$ ranging from 0.8$-$1.2, but with two exceptions for which {\it P}$_2$ is much longer and the ratio is $\sim$2.
In Table~\ref{results}, we summarize the results of our period and light-curve study of these 12 objects.  
Included are the maximum change in {\it V} brightness in a season ($\Delta${\it V}) based on the VUO-new and ASAS-SN light curves, the effective temperature based upon a high-resolution spectroscopic analysis ({\it T}$_{\rm eff}$), {\it P}$_1$ and  {\it P}$_2$ from this study, the period ratio ({\it P}$_2$/{\it P}$_1$), the period found from our previous study \citep[{\it P}$_{\rm old}$;][]{hri10}, the period found in this study from the combination of the VUO-new, VUO-old, and ASAS-SN {\it V} light curves from 1994$-$2018 ({\it P}$_{\rm all}$), and brief comments on any long-term trends in the overall brightness. 
We have also included the same properties for five additional C-rich PPNe from the Milky Way Galaxy for which we have recently determined periods.  They have no earlier ($<$2009) period studies. 

\placetable{results}

\subsection{Period-Temperature Relationship and Investigating the Evidence for Period Change with Time}

\begin{figure}\figurenum{8}\epsscale{1.1} 
\plotone{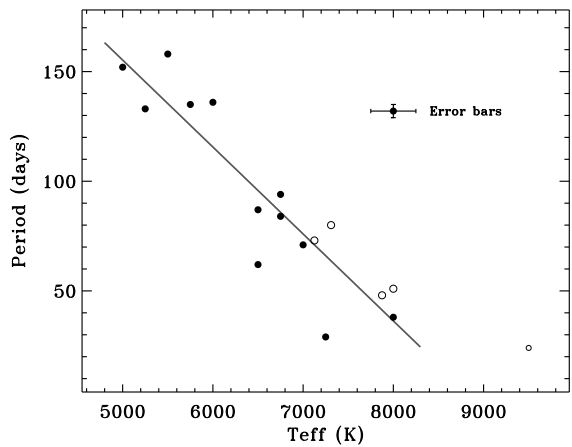}
\caption{Plot showing the relationship between the pulsation periods and the effective temperatures of the stars.
A clear inverse, linear relationship is found.  
The stars from this present study are shown as filled circles, those from other studies as open circles, and the one with a less-certain period shown with a smaller symbol, as listed in Table~\ref{results}.
The solid line in the bottom panel is a least-squares fit to the data between 5000 and 8000 K.
Conservative average error bars of $\pm$3 days and $\pm$250 K are shown.
\label{PT_plot}}
\epsscale{1.0}
\end{figure}

We found in our initial study of C-rich PPNe that there exists a clear relationship between the period ({\it P}) and effective temperature ({\it T}$_{\rm eff}$).
This relationship is shown in Figure~\ref{PT_plot}, where we have plotted the newer {\it P} values from this study and also included data from all additional C-rich PPNe in the Milky Way Galaxy for which reliable values are known.
The relationship is approximately linear from 5000 to 8000 K, with a slope of $-$0.040 $\pm$ 0.005 day~K$^{-1}$, but has a much shallower slope thereafter.\footnote{Although there is only one additional object, of less-certain period, with temperature greater than 8000 K, 
the linear relationship cannot continue to a temperature much higher than 8000 K before reaching {\it P} = 0 days.}
A study of a smaller sample of six PPNe in the Magellanic Clouds shows that they also fit well with this relationship \citep{hri15a}.  
However, we have not included these, as we want to present a more homogenous sample, and the objects in the Magellanic Clouds possess a lower metallicity.

\placefigure{PT_plot}

Since post-AGB stars increase in temperature as they evolve at approximately constant luminosity across the H-R diagram, the above slope can be used to find the approximate rate at which the period of a typical carbon-rich, post-AGB star would be expected to decrease with time during this temperature interval.
This raises the following interesting question: 
{\it Can we expect to see this period change (evolution) in real time for an individual star?}

To investigate this, one needs to know the rate of evolution of a post-AGB star's effective temperature with time.  This is predicted in models of post-AGB evolution.
In our earlier study, we used the models of \citet{blo95} and the analysis by \citet{stef98}.  This latter is based on a star of  {\it M}$_{\rm init}$ = 3.0 M$_{\sun}$ evolving from {\it T}$_{\rm eff}$ = 6050 to 8500 K, from which we determined an average rate on 3.6 K~yr$^{-1}$.
We now have available newer models incorporating better physics published by \citet{milb16}.  
Miller Bertolami divided the post-AGB phase into two parts, 
(1) the initial {\it transition} interval from the end of the AGB (defined as the point when  {\it M}$_{\rm env}$ decreases to 0.01 {\it M}$_{\rm star}$), 
when mass loss from winds might still be important, 
and (2) the faster evolving {\it crossing} interval of the post-AGB phase, defined to begin at log {\it T}$_{\rm eff}$ = 3.85 ({\it T}$_{\rm eff}$=7100 K), 
when mass loss by winds is not important, and the evolution speed depends almost entirely on the H-burning layer.
Based on his model with {\it M}$_{\rm init}$ = 2.0 M$_{\sun}$ and metallicity {\it Z} = 0.01, this leads to an average rate of 1.3 K~yr$^{-1}$ during this transition phase.
However, this is an average value, and the actual rate depends upon the mass-loss rate and the corresponding temperature of the star, which was described in more detail by \citet{milb19}.  There one finds that the heating rate d{\it T$_{\rm eff}$}/d{\it t} increases approximately exponentially from {\it T}$_{\rm eff}$ = 4000 up to about {\it T}$_{\rm eff}$ = 8000$-$10,000 K, depending upon the final (or initial) mass, and then remains approximately constant at that higher rate across to the start of the PN phase ({\it T}$_{\rm eff}$ $\approx$ 30,000 K).
Thus the rate of temperature increase is low when the star first evolves from the AGB to the post-AGB phase.
For  {\it M}$_{\rm init}$ = 2.0 M$_{\sun}$, the rate d{\it T}$_{\rm eff}$/d{\it t} increases from 4.0 to 36 to 53 K~yr$^{-1}$ as {\it T}$_{\rm eff}$ increases from 5500 to 7000 to 8000 K, respectively.  
Using the slope of the {\it P}$-${\it T} relationship, this leads to predicted period change rates of $-$0.16 to $-$1.4 to $-$2.1 day~yr$^{-1}$, respectively.  The rates are correspondingly lower (higher) for stars with {\it M}$_{\rm init}$ = 1.5 (2.5) M$_{\sun}$, respectively.
The predicted period change rates for {\it M}$_{\rm init}$ = 1.5, 2.0, 2.5, and 3.0 M$_{\sun}$ are listed in more detail in Table~\ref{P-rates}, 
where we list the heating rate (d{\it T}$_{\rm eff}$/d{\it t}), the rate of period change (d{\it P}/d{\it t}), and the resulting period change in a decade. 

These masses were chosen since C-rich stars are found to form from stars within a limited range of initial masses, $\sim$1.5$-$3.5 M$_{\sun}$ \citep{slo12,slo16}.
The values for {\it M}$_{\rm init}$ = 1.0 and 1.25 M$_{\sun}$ were also included to illustrate the much slower heating rates of lower-mass stars.
Reviewing the results for carbon star masses suggests that, over a decade, the predicted decrease in period for {\it T$_{\rm eff}$} = 5500 K would amount to about 2 days, 
while for {\it T$_{\rm eff}$} = 7000 K, the predicted decrease is on the order of 10$-$20 days, and even twice as large at {\it T$_{\rm eff}$} = 8000 K.
Thus, in principle they should be measurable.

We can compare the results of our earlier study (1994$-$2007) with these more recent results (2008-2018) to investigate evidence for period changes over that time interval. 
In Table~\ref{results} are listed the periods found from our earlier study ({\it P}$_{\rm old}$). 
It must be recognized from the outset that the light curves are complicated, and there appear to be multiple periods present that complicate the interpretation.  
With that in mind, we examine the results for the 11 PPNe, excluding IRAS 20000+3239, which did not have a VUO-new light curve. 
For those (four) with longer periods (130$-$160 days) and lower temperatures (5000$-$6000 K), we find period differences ranging from 0 to +5 days.  Given the uncertainties, we consider these to be relatively similar results between the studies of the two epochs. 
For those (five) with medium-length periods (60$-$100 days) and and somewhat higher temperatures (6500$-$7000 K), we find little change for four (0 to $-$2 days), while for AFGL 2688, we have earlier mentioned the uncertainty in the older period, and so we disregard its value. 
For the two remaining, shorter-period (29$-$39 days) and hotter (7250$-$8000 K) objects, we find evidence for significant period changes.  The period for IRAS 07134+1005 has changed from $\sim$39 days from 1994$-$2007 to 29 days in the present (2008$-$2018), a decrease of 10 days. 
For IRAS 19500$-$1709, the periods found in the present 2008$-$2018 and the older (1994$-$2007) studies are both 38 days.  However, the period found in the more recent (2018$-$2021) ASAS-SN {\it g} data is 29 days, possibly suggesting a very recent decrease.  This, of course, needs to be confirmed by additional monitoring of the star.
Although there are a lot of uncertainties in the observational results due to the multiple periods, varying amplitudes, and nonperiodic physical phenomenon (such as shocks) contributing to the light variations, one main result appears to emerge $-$ that one or both of the hotter, shorter-period objects has shown a significant period decrease over the past 25 yr.
Comparison with the models of \citet{milb19} indicates that for a star of {\it T}$_{\rm eff}$ = 7000 K and {\it M}$_{\rm init}$ = 1.5$-$2.0 {\it M}$_{\sun}$, a change of 10 days can occur in 10 years.
At the higher value of {\it T}$_{\rm eff}$ = 8000 K and {\it M}$_{\rm init}$ = 1.5$-$2.0 {\it M}$_{\sun}$, such a change can occur in about half that time.
These are very preliminary results for the change in period and consequently in temperature of a post-AGB star with time. 
However, they do appear to indicate the correct order of magnitude for the model heating rates for stars at the lower mass boundary for carbon stars. 
They are certainly too preliminary to use to confirm the heating rate or the {\it M}$_{\rm final}$ of the star, but they may be a first step in this direction. 

There are other aspects of the evolutionary models such as those of \citet{milb16,milb19} that can be compared with observations.
A more direct one is that the observed effective temperature is predicted to increase with time.  
We have referred to this above when we calculated the predicted rate for period change.  As mentioned above, this rate increases with temperature and with the mass of the star (see Table~\ref{P-rates}).  
A star of {\it T}$_{\rm eff}$ = 5500 K and {\it M}$_{\rm init}$ = 1.5$-$2.0 {\it M}$_{\sun}$ changes little in temperature in a decade, while one of {\it T}$_{\rm eff}$ = 8000 K and {\it M}$_{\rm init}$ = 1.5$-$2.0 {\it M}$_{\sun}$ changes by 400$-$500 K in a decade.  While this is measurable, it is complicated by the fact that the stars change in temperature during the pulsation cycle, being hottest when brightest and smallest \citep{hri13}.
For the stars in the present study, temperature changes due to pulsation of 300$-$700 K have been determined based upon color changes \citep{hri10}.  
In a recent spectroscopic study, \citet{puk22} attempted to determine the evolution rate of the temperature of the oxygen-rich PPN HD 161796 (IRAS 17436+5003) over an 18 yr interval, but they encountered this very problem.
Thus, it would likely require a high-resolution spectroscopic study plus a contemporaneous light-curve monitoring program to disentangle the two effects.  But this could be done, particularly for the hotter two objects, which are bright ({\it V}=8$-$9 mag), by observing them in epochs separated by a decade or more.  And such spectroscopically determined temperatures would not be compromised by changes in the line-of-sight dust opacity, which can effect the colors.

\subsection{Variability Amplitude and Temperature Relationship}

\begin{figure}\figurenum{9}\epsscale{1.1} 
\plotone{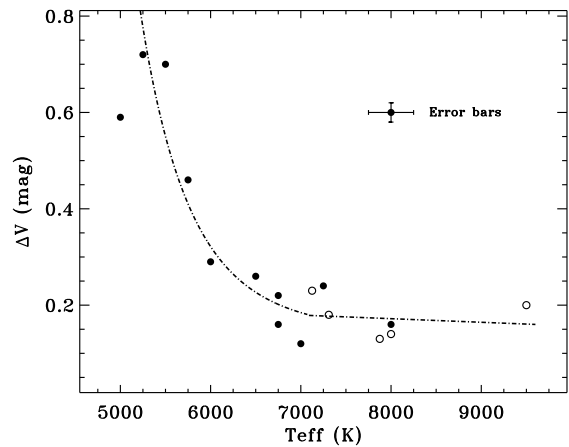}
\caption{Plot showing the relationship between the maximum brightness change in a season ($\Delta$V) and the effective temperature ({\it T}$_{\rm eff}$) of the stars.
The values are much larger at the cooler temperatures and then decrease rapidly by {\it T}$_{\rm eff}$ = 6500$-$7000 K.
The stars from the present study are shown as filled circles, and those from other studies as open circles, as listed in Table~\ref{results}.
AFGL 2688 is not included because the measured variation in starlight is compromised by the large scattering area of the lobe.
The dotted-dashed line is a free-hand representation of the trend in the data.
Estimated average error bars of $\pm$0.02 mag and $\pm$250 K are shown.
\label{delV-T_plot}}
\epsscale{1.0}
\end{figure}

As has been documented previously \citep{hri10}, the peak-to-peak variation in brightness of the stars is correlated to the temperature of the stars.  
In Table~\ref{results} we list for each of the program stars the maximum change in {\it V} observed in a season ($\Delta${\it V}), based on the combined VUO-new and ASAS-SN light curve. 
The cooler stars ({\it T}$_{\rm eff}$=5000$-$5750 K) have larger variations (0.5$-$0.7 mag) than the higher-temperature stars.
This is shown in Figure~\ref{delV-T_plot}.
At {\it T}$_{\rm eff}$ $\approx$ 6500$-$7000 K, the maximum amplitude has decreased to an approximately constant lower value $\sim$0.2 mag.
This rapid drop-off  in amplitude may be related to the ongoing mass loss in the post-AGB phase.
The evolutionary models include a decreasing rate of mass loss during the initial portion of the post-AGB evolution \citep{vas94,blo95,milb16}, which effectively terminates at some temperature.  The \citet{milb16} models have this occur at log {\it T}$_{\rm eff}$ = 3.85 ({\it T}$_{\rm eff}$$\simeq$7100 K), similar to the temperature inflection point that can be seen in the $\Delta${\it V}$-${\it T}$_{\rm eff}$ plot.  

\placefigure{delV-T_plot}

\subsection{Long-term Trends in Brightness}
\label{trends}

The presence of long-term brightness trends in the objects can be examined by combining the VUO-old and VUO-new data.  This was shown visually for 10 of the objects in Figure~\ref{fig4} and for IRAS 04296+3429 and 23304+6147 by \citet[Fig. 2]{hri21}.
General, monotonic increases in {\it V} brightness are seen in IRAS 05341+1347 ($-$0.15 mag), AFGL 2688 ($-$0.30 mag), and IRAS 04296+3429 ($-$0.17 mag).
IRAS Z02229+6208 and 07430+1115 show more complicated increases in brightness of approximately $-$0.3 and $-$0.2 mag, respectively.  
In contrast, IRAS 22223+4327 shows a decrease in brightness (+0.18 mag) that appears to have settled at a constant level since 2009.
The brightness increase in AFGL 2688 has been documented since 1920 \citep{got76}.
As mentioned earlier, IRAS 19500$-$1709 dropped in brightness by 0.12 mag ({\it V}) in 1999, gradually rose to a level $-$0.30 mag brighter than the original brightness level by 2012, and has subsequently decreased by 0.07 mag.
There is a suggestion that a similar event but of larger magnitude occurred in IRAS 05113+1347, which decreased in brightness by $\sim$0.6 mag in 2011 and has been slowly increasing since then.
As mentioned when we discussed the individual objects, most appear to show some smaller changes in the median level from season to season.
We attribute these primarily to changes in the circumstellar dust opacity.  The sudden drops can be attributed to recent mass ejection events or the passing of a circumstellar dust cloud through the line of sight.  

The only object in this study for which we have evidence of a longer-term periodic behavior is IRAS 22272+5435.  This object had a significant period of 1.8 yr in the VUO-new and the combined VUO-new and ASAS-SN data.  When we also combined the VUO-old data, this period was not significant, but a longer period of 3.2 yr emerged.  This object is deserving of further study. 
No long-term periodic behavior is seen in any of the other light curves that might be evidence of a binary companion or motion within a circumbinary disk.  However, we cannot rule out periods on the order of twice the observing interval or longer, particularly in cases like IRAS 05341+0852 in which the variations can be fit by half of a long-period sine curve.
Long-term (5$-$19 yr)  periodic behavior has been seen in several other PPNe and post-AGB objects, including two of the C-rich PPNe we recently studied \citep{hri20a} and IRAS 08005$-$2356 \citep{man21}.

A main reason for the interest in binarity is as a shaping mechanism for the surrounding nebulae.  Many PPNe have a bipolar morphology, and shaping by the interaction of a close binary is suggested as the cause.  Evidence for binary companions has been lacking.  
\citet{hri17} carried out long-term radial-velocity monitoring of a sample of seven of the brightest PPNe without finding any confirmed binaries.
However, recently a combined radial-velocity and photometric monitoring study has discovered the first certain binary in the PPNe IRAS 08005$-$2356 \citep[V510 Pup;][]{man21}.  
Another frequently cited candidate is HD 101584 \citep[see][and references therein]{jon20}.

\section{Summary and Conclusions}
\label{summary}

We carried out photometric monitoring from 2008$-$2018 of 11 carbon-rich PPNe of F$-$G spectral type.  These data, in combination with recent ASAS-SN data, were analyzed for periodicity.
These new results were then compared with the results of our earlier (1994$-$2007) monitoring study of these same objects plus one additional PPNe. 
The main results of this study are listed below.

1. All 12 objects were found to vary periodically, with periods ranging from 158 to 29 days.  These periods are generally in accord with those found in the earlier study.  

2. As found previously, the objects display an approximately inverse linear relationship between {\it T}$_{\rm eff}$ and period, with a slope of $-$0.040 day~K$^{-1}$.

3. There is evidence in one object for a decrease in period of 9 days between the two data sets, and possible evidence for a change in a second object.  These are the two highest-temperature objects in the sample (7250 and 8000 K).

4. Such a period change is consistent with the evolution expected for a low-mass carbon star ({\it M}$_{\rm init}$ 1.5$-$2 M$_{\sun}$) of such a temperature, based on the above slope and the stellar evolution models of \citet{milb16,milb19}. 

5. Secondary periods are seen in most of the objects, with ratios {\it P}$_2$$/${\it P}$_1$  ranging from 0.8$-$1.2 except in two cases, where the ratio is $\sim$2. 

6. Long-term changes in brightness are seen in seven of the objects, based on the combined data sets from 1994$-$2018, with six of them increasing (by $-$0.15 to $-$0.30 mag) and one of them decreasing (by 0.12 mag) over that time interval.  We attribute this to changes in the opacity of the circumstellar dust.

7. There is a suggestion in one object, IRAS 22272+5435, of a cyclical variation in brightness with a period of $\sim$1.8 yr period, which could possibly be caused by an orbiting companion.  

The evidence for period changes over the order of a decade in one or two objects and the evidence for a possible companion-induced periodic light variation of $\sim$2 yr in another object are all of great astronomical interest as we seek to better understand the evolution of these transitional objects.  
Detailed follow-up observations should be made of these three objects to confirm and refine these results.  
Also, second-epoch observations should be made of the several additional PPNe with short periods and high temperatures (7875$-$9500 K) listed in the lower portion of Table 5, to see if they also show evidence of period changes.
Future programs combining the results from light-curve monitoring with contemporaneous high-resolution spectroscopic studies would provide a more direct way to investigate such evolution; however, these would require more telescope time.
Continued photometric monitoring of the brightness levels of all of these PPNe would be useful to document overall brightness changes.

\acknowledgments
We acknowledge with thanks the many Valparaiso University undergraduate students who carried out the photometric observations at the Valparaiso University Observatory from 2008 to 2018.
These are, in chronological order, Ryan McGuire, Sam Schaub, Chris Wagner, Kristie (Shaw) Nault, Zach Nault, Wesley Cheek, Joel Rogers, Rachael (Jensema) Filwett, Chris Miko, Austin Bain, Hannah Rotter, Aaron Seider, Allyse Appel, Brendan Ferris, Justin Reed, Jacob Bowman, Ryan Braun, Dani Crispo, Stephen Freund,  Chris Morrissey, Cole Hancock, Abigail Vance, Kathryn Willenbrink, Matthew Bremer, Avery Jackson, David Vogl, Tim Bimler, and Sammantha Nowak-Wolff.
We thank Marcelo Miller Bertolami for kindly providing graphs and tables of smoothed heating tracks for his models, and we thank Todd Hillwig and the anonymous referee for helpful comments.
The ongoing technical support of Paul Nord is gratefully acknowledged.  Matthew Bremer assisted in an earlier analysis of some of these objects.
B.J.H. acknowledges ongoing support from the National Science Foundation (1413660) and the Indiana Space Grant Consortium.
This research has made use of the SIMBAD and VizieR databases, operated at CDS, Strasbourg,
France, and NASA's Astrophysical Data System.

\facility{ASAS-SN)}

\clearpage

\tablenum{1}
\begin{deluxetable}{crrrrrrrccl}
\tablecolumns{11} \tabletypesize{\scriptsize}
\tablecaption{List of C-Rich PPNs Observed\label{object_list}}
\rotate
\tabletypesize{\footnotesize} 
\tablewidth{0pt} \tablehead{
\colhead{IRAS ID}&\colhead{2MASS ID}&\colhead{Gaia ID}&\colhead{R.A.\tablenotemark{a}}&\colhead{Decl.\tablenotemark{a}}
&\colhead{{\it l}}&\colhead{{\it b}}
&\colhead{{\it V}\tablenotemark{b}}&\colhead{{\it B$-$V}\tablenotemark{b}}&\colhead{Sp.T.}&\colhead{Other ID} \\
&&&(2000.0)&(2000.0)&($\arcdeg$)&($\arcdeg$)&\colhead{(mag)} &\colhead{(mag)} & & } 
\startdata
Z02229+6208\tablenotemark{c} &02264179+6221219 & 513671461473684352 & 02:26:41.8 & +62:21:22 & 133.7 & +1.5 & 12.1 & 2.8 & G8-K0 0-Ia:& \nodata \\
04296+3429 & 04325697+3436123 & 173086700992466688&04:32:57.0 & +34:36:12 & 166.2 & $-$9.0 & 14.2 & 2.0 & G0 Ia & \nodata \\
05113+1347 & 05140775+1350282 & 3388902129107252992 &05:14:07.8 & +13:50:28 & 188.9 & $-$14.3 & 12.4 & 2.1 & G8 Ia & \nodata \\
05341+0852 & 05365506+0854087 & 3334854780347915520 &05:36:55.1 & +08:54:09 & 196.2 & $-$12.1 & 13.6 & 1.8 & G2 0-Ia: & \nodata \\
07134+1005 & 07161025+0959480 & 3156171118495247360 &07:16:10.3 & +09:59:48 & 206.7 & +10.0 & 8.2 & 0.9 & F5 I & HD~56126, CY CMi \\
07430+1115 & 07455139+1108196 & 3151417586128916864 &07:45:51.4 & +11:08:19 & 208.9 & +17.1 & 12.6 & 1.9 & G5 0-Ia: & \nodata \\
19500$-$1709 & 19525269$-$1701503 & 6871175064823382912 &19:52:52.7 & $-$17:01:50 & 24.0 & $-$21.0 & 8.7 & 0.5 & F3 I & HD~187885, V5122 Sgr \\
20000+3239 & 20015951+3247328  & 2034134414507432064 & 20:01:59.5 & +32:47:33 & 69.7 & +1.2 & 13.3 & 2.7 & G8 Ia & \nodata \\
22223+4327 &22243142+4343109 & 1958757291756223104 & 22:24:31.4 & +43:43:11 & 96.8 & $-$11.6 & 9.7 & 0.9 & G0 Ia & V448 Lac \\ 
22272+5435 &J22291039+5451062 & 2006425553228658816 & 22:29:10.4 & +54:51:06 & 103.3 & $-$2.5 & 9.0 & 2.0 & G5 Ia & HD~235858, V354~Lac \\
23304+6147 &J23324479+6203491 & 2015785313459952128 & 23:32:44.8 & +62:03:49 & 113.9 & +0.6 & 13.1 & 2.3 & G2 Ia & \nodata \\
AFGL~2688\tablenotemark{d} &J21021878+3641412 & \nodata\tablenotemark{e} &21:02:18.8 & +36:41:41 & 80.2 & $-$6.5 & 12.2 & \nodata & F5~Iae & Egg~Nebula, V1610~Cyg \\
\enddata
\tablenotetext{a}{Coordinates from the 2MASS Catalog.}
\tablenotetext{b}{These values are all variable as discussed in this paper.  They are based on our measurements, except for the {\it B}$-${\it V} measurement of IRAS 07134$+$1005 and the {\it V} measurement of AFGL 2688, which are from SIMBAD. }
\tablenotetext{c}{${\rm L}$isted inisted in the {\it IRAS} Faint Source Reject File (thus the ``Z'') but not in the Point Source Catalog. }
\tablenotetext{d}{Not included in the {\it IRAS} Point Source Catalog. }
\tablenotetext{e}{Not included in the Gaia Catalog. } 
\end{deluxetable}

\clearpage

\tablenum{2}
\begin{deluxetable}{rlrrrrcrrr}
\tablecaption{Primary Comparison Star Identifications, Standard Magnitudes and Colors, Along with Statistics of the PPN Light Curves \label{std_comp}}
\tabletypesize{\footnotesize}
\tablewidth{0pt} \tablehead{\colhead{IRAS Field}
&\colhead{Star} &\colhead{2MASS ID} &\colhead{{\it V}} &\colhead{{\it B$-$V}} 
&\colhead{{\it V$-$R}$_C$} & \colhead{Ref.\tablenotemark{a}} &\colhead{no.({\it V})\tablenotemark{b}} &\colhead{no.({\it R}$_C$)\tablenotemark{b}} &\colhead{no.({\it B})\tablenotemark{b}}  \\
\colhead{} &\colhead{} &\colhead{}  &\colhead{(mag)} &\colhead{(mag)} &\colhead{(mag)}
&\colhead{} &\colhead{} &\colhead{} &\colhead{}} \startdata
Z02229$+$6208 & C1\tablenotemark{c} & 02270480+6221258 & 10.85 &\nodata & 0.40 & 1 & 146 &155 & 0 \\
04296+3429 & C1 & 04331115+3436511 & 13.18 & \nodata & 0.50 & 2 & 67 & 93 & 0 \\
05113+1347 & C1 & 05142163+1355114 & 12.30 & 0.84 & \nodata & 3 & 53 & 62 & 0 \\
05341+0852 & C1\tablenotemark{d,e} & 05365435+0855346 & 11.82 &\nodata & \nodata  & 1 & 41 & 47 & 0 \\
07134+1005 & C1 & 07161930+0959483 & 10.84 & 0.98 & \nodata & 3 & 164 & 167 & 143 \\
07430+1115 & C1\tablenotemark{d} & 07454334+1109112 & 11.97 & 0.45 & 0.28 & 1 & 94 & 101 & 0 \\
19500$-$1709 & C1\tablenotemark{f} & 19525346$-$1659560 & 10.07 & 0.44 & 0.25  & 4 & 293 & 300 & 247 \\
20000+3239 & C1 & 20014554+3247569 & 11.43 &\nodata & 0.60 & 1 & 0 & 0 & 0 \\
22223+4327 & C1 & 22241785+4342556 & 11.09 & 0.54 & 0.32 & 4 & 389 & 394 & 339 \\
22272+5435 & C1\tablenotemark{g} & 22285036+5452299 & 11.16 & 1.11 & 0.60 & 4 & 393 & 395 & 346 \\
23304+6147 & C1\tablenotemark{h} & 23324587+6203175 & 12.73 & \nodata & 0.46 & 1 & 192 & 197 & 0 \\
AFGL~2688 & C1 & 21021186+3642044 & 12.20 &\nodata & 0.74 & 1 & 216 & 234 & 0 \\
\enddata
\tablecomments{Uncertainties in the brightness and color are $\pm$0.01$-$0.02 mag, except for reference 2, where they are $\pm$0.02$-$0.04 mag.}
\tablenotetext{a}{References to the photometry are as follows: (1) \citet{hri10}, (2) \citet{hri20a}, (3) \citet[APASS]{hen12}, (4) this study.} 
\tablenotetext{b}{Represents the number of new VUO observations made of each of the PPNe fields.} 
\tablenotetext{c}{There is a large difference between our measurements of this comparison star and those of \citet{hen12} $-$ {\it V} = 10.69, ({\it B$-$V}) = 0.51.  Our differential observations of the star from 2008$-$2018 do not indicate a variation of the star at larger than the level of $\pm$0.02 mag.  Thus, presumably one of these standardized measurements is in error. }
\tablenotetext{d}{C1 changed from \citet{hri10}.  To combine the older data set with the present, the following offsets need to be added to the 2010 data: IRAS 05341+0852, offsets are $-$0.949 ({\it V}) and $-$0.846 ({\it R}$_C$); IRAS 07430+1115, offsets are $-$1.456 ({\it V}) and $-$1.626 ({\it R}$_C$). 
(In addition, we note that the comparison stars for IRAS 05341+0852 were incorrectly identified by \citet[Table 4]{hri10}. 
The one identified in that study as C1 was actually C3, C2 was actually C1, and C3 was actually C2 in that study.  
The C1 used in that study was actually 2MASS 05365669+0852563, {\it V}=10.8.) }
\tablenotetext{e}{Also observed by \citet{hen12} $-$ {\it V} = 11.80, ({\it B$-$V}) = 0.67.  }
\tablenotetext{f}{Note correction to \citet[Table 4]{hri10}, where the IDs for C1 and C2 were mistakenly reversed, although the magnitudes listed for C1 and C2 are correct.} 
\tablenotetext{g}{The comparison stars are the same as used by \citet{hri10} but differ from those used by \citet{hri13}, with present C1 = C3 in \citet{hri13}.} 
\tablenotetext{h}{The comparison stars are the same as used by \citet{hri10,hri20a} but the values of {\it V$-$R}$_C$ for C1 and C3 were mistakenly reversed by \citet[Table 3]{hri20a}.} 
\end{deluxetable}

\clearpage

\tablenum{3}
\begin{deluxetable}{rcrcrcr}
\tablecolumns{7} \tabletypesize{\scriptsize}
\tablecaption{Differential Standard {\it BVR$_{\rm C}$} Magnitudes for the Program Objects  from 2008$-$2018 
 \label{std_diffmag}}  
 \tablewidth{0pt} \tablehead{\colhead{Object Name} & \colhead{HJD $-$ 2,400,000} & \colhead{$\Delta${\it V}}  
& \colhead{HJD $-$ 2,400,000} & \colhead{$\Delta${\it R}$_{\rm C}$} 
& \colhead{HJD $-$ 2,400,000} & \colhead{$\Delta${\it B}}}
\startdata
IRAS 22223+5435 & \nodata        &\nodata & 54654.7279 & -1.433 & \nodata       & \nodata \\
IRAS 22223+5435 & \nodata        &\nodata & 54657.8651 & -1.439 & \nodata       & \nodata \\
IRAS 22223+5435 & 54662.7959 & -1.197 & 54662.7942 & -1.431 & 54662.8042 & -0.775 \\
IRAS 22223+5435 & 54663.8319 & -1.201 & 54663.8306 & -1.436 & 54663.8341 & -0.789 \\
IRAS 22223+5435 & 54664.7696 & -1.200 & 54664.7676 & -1.431 & 54664.7722 & -0.786 \\
IRAS 22223+5435 & 54665.6989 & -1.187 & 54665.6975 & -1.416 & 54665.7023 & -0.784 \\
IRAS 22223+5435 & 54670.7185 & -1.203 & 54670.7168 & -1.437 & 54670.7460 & -0.792 \\
IRAS 22223+5435 & 54672.7058 & -1.202 & 54672.7048 & -1.437 & 54672.7083 & -0.779 \\
\enddata
\tablecomments{This table is available in its entirety in machine-readable form. }
\end{deluxetable}

\clearpage

\tablenum{4}
\begin{deluxetable}{lcrrrrrrrrrrrrr}
\tablecolumns{15} \tabletypesize{\scriptsize}
\tablecaption{Main Results of the Periodogram Study of the Light Curves\tablenotemark{a,b}\label{periods}}
\rotate
\tabletypesize{\footnotesize} 
\tablewidth{0pt} \tablehead{ 
\colhead{IRAS ID} &\colhead{Filter} & \colhead{Years} & \colhead{Data\tablenotemark{c}} & \colhead{No.} & \colhead{{\it P}$_1$} & \colhead{{\it A}$_1$} & \colhead{{\it $\phi$}$_1$\tablenotemark{d}} & \colhead{{\it P}$_2$}&\colhead{{\it A}$_2$} &\colhead{$\phi$$_2$\tablenotemark{d}} & \colhead{{\it P}$_3$} & \colhead{{\it A}$_3$} & \colhead{$\phi$$_3$\tablenotemark{d}} & \colhead{$\sigma$\tablenotemark{e}} \\
 & & & \colhead{Sets} &\colhead{Obs.} & \colhead{(days)}&\colhead{(mag)} & &\colhead{(days)}&\colhead{(mag)}
 & &\colhead{(days)} &\colhead{(mag)}&  &\colhead{(mag)} }
\startdata
Z02229$+$6208 & {\it V} & 2009-2018 & 1,2  & 337 & 157.8 & 0.073 & 0.80 & 261.2 & 0.067 & 0.031 & \nodata & \nodata & \nodata  & 0.103 \\
04296$+$3429 & {\it V} & 2008-2019 & 1,2 & 187 & 71.2 & 0.024 & 0.99 & 68.8 & 0.017 & 0.47 & 76.7 & 0.016 & 0.87 & 0.020 \\
05113$+$1347 & {\it V} & 2010-2018 & 1,2 & 367 & 132.8 & 0.160 & 0.40 & 121.4 & 0.114 & 0.97 & 135.5 & 0.166 & 0.04 & 0.080 \\
05341$+$0852 & {\it V} & 2009-2018 & 1,2 & 179 &   94.1 & 0.045 & 0.37 & 74.6 & 0.024 & 0.77 & \nodata & \nodata & \nodata  & 0.026 \\
07134$+$1005 & {\it V} & 2009-2018 & 1,2 & 358 &  29.4 & 0.021 & 0.74 & 36.5 & 0.017 & 0.91 & 58.0 & 0.017 & 0.02  & 0.035 \\
 & {\it g} & 2018-2021 & 2 & 189 &  29.5 & 0.088 & 0.80 & 14.8 & 0.043 & 0.63 & \nodata & \nodata & \nodata & 0.059 \\
07430$+$1115 & {\it V} & 2009-2018 & 1,2 & 371 & 135.9 & 0.048 & 0.36 & 298.2 & 0.051 & 0.02 & \nodata & \nodata & \nodata  & 0.046 \\
19500$-$1709 & {\it V} & 2009-2017 & 1 & 289 & 38.2 & 0.019 & 0.85 & \nodata & \nodata & \nodata & \nodata & \nodata & \nodata & 0.030 \\
 & {\it g} & 2018-2020  & 2 & 381 & 29.5 & 0.069 & 0.72 & 14.7 & 0.038 & 0.16 & \nodata & \nodata & \nodata & 0.076 \\
20000$+$3239 & {\it V} & 1994-2017 & 2,3 & 340 & 152.5 & 0.075 & 0.66 & 130.4 & 0.050 & 0.88 & \nodata & \nodata & \nodata & 0.075 \\
22223$+$4327 & {\it V} & 2008-2019 & 1,2 & 535 &  86.6 & 0.051 & 0.33 &  91.2 & 0.021 & 0.29 & 116.1 & 0.017 & 0.31  & 0.028 \\
22272$+$5435 & {\it V} & 2008-2019 & 1,2 & 614 & 134.8 & 0.055 & 0.04 & 121.6 & 0.045 & 0.73 & 154.0 & 0.045 & 0.62  & 0.067 \\
23304$+$6147 & {\it V} & 2008-2019 & 1,2 & 394 & 83.6 & 0.027 & 0.69 & 87.3 & 0.022 & 0.91 & \nodata & \nodata & \nodata & 0.031 \\
AFGL~2688 & {\it V} & 2009-2018 & 1,2 & 600 & 62.0 & 0.015 & 0.95 & 57.6 & 0.016 & 0.41 & \nodata & \nodata & \nodata  & 0.026 \\
\enddata
\tablenotetext{a}{The uncertainties in the parameters are as follows: period ({\it P}) $-$ $\pm$0.1$-$0.3 days for pulsation, except for the two with P$\sim$250$-$300 days with uncertainty of $\pm$1.0 day; amplitude ({\it A}) $-$  $\pm$0.002$-$0.007 mag; phase ($\phi$)  $-$ $\pm$0.01$-$0.02. }
\tablenotetext{b}{Analysis based on the seasonally normalized light curve, with trends removed.}
\tablenotetext{c}{1 = VUO-new, 2 = ASAS-SN, 3 = VUO-old.}
\tablenotetext{d}{The phases are determined based on the epoch of 2,455,600.0000, and they each represent the phase derived from a sine-curve fit to the data, not the phase of minimum light.}
\tablenotetext{e}{Standard deviation of the observations from the sine-curve fit.}
\end{deluxetable}

\clearpage

\tablenum{5}
\begin{deluxetable}{rrrcrccrrrl}
\tablecaption{Results of Our Period and Light-curve Study\label{results}} 
\tabletypesize{\footnotesize}
\tablewidth{0pt} \tablehead{ \colhead{IRAS ID} &\colhead{{\it V}} &\colhead{$\Delta${\it V}\tablenotemark{a}} 
&\colhead{SpT} &\colhead{{\it T}$_{\rm eff}$\tablenotemark{b}} &\colhead{{\it P}$_1$} &\colhead{{\it P}$_2$} &\colhead{{\it P}$_2$/{\it P}$_1$} &\colhead{{\it P}$_{\rm old}$\tablenotemark{c}} &\colhead{{\it P}$_{\rm 1,all}$\tablenotemark{d}}  & \colhead{Trend\tablenotemark{e}}\\
\colhead{} &\colhead{(mag)} &\colhead{(mag)} &\colhead{} &\colhead{(K)} &\colhead{(days)} &\colhead{(days)} &\colhead{} &\colhead{(days)} & \colhead{(days)} & \colhead{}}
\startdata
Z02229$+$6208 &  12.1 & 0.70 & G8-K0 0-Ia: & 5500 & 158 & 261& 1.66 & 153 & 155 & Inc \\
04296$+$3429 &  14.2 & 0.12 & G0 Ia  & 7000 & 71 & 69 & 0.96 & 71: & 71 & Inc  \\
05113$+$1347  & 12.4 & 0.72 & G8 Ia & 5250 & 133 & 121 & 0.91 & 133 & 122,137 & Drop; Inc\\
05341$+$0852 &  13.6 & 0.16 & G2 0-Ia: & 6750 & 94 & 75 & 0.79 & 94 & 94 & Inc \\
07134$+$1005 &   8.2 & 0.24 & F5 I  & 7250 & 29 & 36 & 1.24 & $\sim$39 & \nodata & \nodata  \\
07430$+$1115 &  12.8 & 0.29 & G5 0-Ia: & 6000 & 136 & 298 & 2.19 & 136 & 136 & Inc \\
19500$-$1709 &    8.7 & 0.16 & F3 I & 8000 & 38\tablenotemark{f} &  \nodata &  \nodata & 38 & 38 & Drop; Inc \\
20000$+$3239 &  13.3 & 0.59 & G8 Ia & 5000 & 154\tablenotemark{g}& 133\tablenotemark{g} & 0.86\tablenotemark{g} & 153 & 152,130\tablenotemark{h} & \nodata \\
22223$+$4327 &    9.7 & 0.26 & G0 Ia & 6500 & 87 & 91 & 1.05 & 89 & 87 & Dec \\
22272$+$5435 &    9.0 & 0.46 & G5 Ia& 5750 & 135 & 122 & 0.90 &130 &132 & \nodata \\
23304$+$6147 &  13.1 & 0.22 & G2 Ia  & 6750 & 84 & 87 & 1.04 & 85 & 84 & \nodata \\
AFGL~2688 &  12.2 & 0.14 & F5~Iae & 6500 & 62 & 58 & 0.93 & $\sim$91: & 62,58 & Inc \\
\tableline 
\multicolumn{10}{c}{Results of Previous Studies\tablenotemark{i}} \\
\tableline 
06530$-$0213 &  14.0 & 0.18 & F5~Ia,G1~I & 7310  & 80 &74 & 0.93 & \nodata & \nodata & Inc \\
08143$-$4406 &  12.3 & 0.23 & F8~I & 7125  & 73 &58 & 0.79 & \nodata & \nodata & P=5.0 yr \\
08281$-$4850 &  13.9 & 0.13 & F0~I & 7875 & 48 & 56 & 1.15 & \nodata & \nodata & \nodata  \\
13245$-$5036 &  12.3 & 0.20 & A7-9~Ie & 9500 & 24: & \nodata & \nodata & \nodata & \nodata & \nodata  \\
14325$-$6428 &  11.8 & 0.14 & F5~I & 8000 & 51 & 47 & 0.91 & \nodata & \nodata & P=18.8 yr \\
\enddata
\tablenotetext{a}{The maximum brightness range observed in a season based on the combined VUO-new and ASAS-SN light curves, except for IRAS 20000+3239, where it is based on the VUO-old light curve}
\tablenotetext{b}{Temperatures from high-resolution spectral observations and abundance analyses: all from \citet{des16} except for Z02229+6208 \citep{red99}, 04296+3429 and 23304+6147 \citep{vanwin00}, 05113+347 \citep{red02}, 20000+3239 \citep{kloch06}, and AFGL 2688 \citep{kloch00}. }
\tablenotetext{c}{From \citet{hri10}. }
\tablenotetext{d}{Based on analysis of combined VUO-old, VUO-new, and ASAS-SN {\it V} light-curve data, except for IRAS 07134+1005 and 19500$-$1709, which did not include the ASAS-SN {\it V} light-curve data.}
\tablenotetext{e}{Brightness trend: Inc = increasing; Dec = decreasing; Drop = sudden drop in brightness; P = value for periodic variation.  There is more quantitative information on these trends in Section~\ref{trends}. }
\tablenotetext{f}{Evidence from the ASAS-SN {\it g} data that the period has decreased to 29.5 days.  However, this needs to be confirmed.}
\tablenotetext{g}{Based on analysis of ASAS-SN {\it V} light curve only, in the absence of VUO-new data.}
\tablenotetext{h}{Based on analysis of combined VUO-old and ASAS-SN {\it V} light curve, but without VUO-new data.}
\tablenotetext{i}{All from \citet{hri21} except IRAS 06530$-$0213 \citep{hri20a}.}
\end{deluxetable}

\clearpage

\tablenum{6}
\begin{deluxetable}{rrrrr}
\tablecaption{Modeled\tablenotemark{a} and Predicted Rates of Temperature and Period Evolution \label{P-rates}} 
\tabletypesize{\footnotesize}
\tablewidth{0pt} \tablehead{ \colhead{{\it T}$_{\rm eff}$} &\colhead{{\it M}$_{\rm init}$} &\colhead{d{\it T}$_{\rm eff}$/d{\it t}\tablenotemark{a}} &\colhead{d{\it P}/d{\it t}\tablenotemark{b}} 
&\colhead{$\Delta${\it P}(10 yr)}  \\
\colhead{} &\colhead{} &\colhead{(K/yr)} &\colhead{(days/yr)} &\colhead{(days)} }
\startdata
5500 & 1.00 & 1.3 & $-$0.05 & $-$0.5  \\
	&  1.25 & 3.0 & $-$0.12  & $-$1.2    \\
        &  1.50 & 3.9 & $-$0.16  & $-$1.6    \\
        & 2.00 & 4.0 & $-$0.16 & $-$1.6   \\
        &  2.50 & 4.9 & $-$-0.20 & $-$2.0    \\
        &  3.00  & 6.4& $-$0.26 & $-$2.6     \\
7000 & 1.00 & 2.7 & $-$0.11 & $-$1.1 \\
	&  1.25 & 8.8 & $-$0.35  & $-$3.5    \\
        &  1.50 & 26 & $-$1.0  & $-$10     \\
        & 2.00 & 36 & $-$1.4 & $-$14   \\
        &  2.50 & 60 & $-$2.4  & $-$24   \\
        &  3.00  & 134 & $-$5.4  & $-$54     \\
8000 & 1.00 & 3.2 & $-$0.13 & $-$1.3 \\
	&  1.25 & 11 & $-$0.44  & $-$4.4    \\
        &  1.5 & 42 & $-$1.7  & $-$17     \\
        & 2.0 & 53 & $-$2.1 & $-$21  \\
        &  2.5 & 129 & $-$5.2 & $-$52   \\
        &  3.0  & 340 & $-$13.6  & $-$136   \\
\enddata
\tablenotetext{a}{Based on modeled heating rate of \cite{milb19} and Miller Bertolami (2022, personal communication) for models with Z = 0.01.}
\tablenotetext{b}{Based on heating rate (Column (3)) and d{\it P}/d{\it T}$_{\rm eff}$ = $-$0.040 determined in this study.  }
\end{deluxetable}

\clearpage

\clearpage

\clearpage

\clearpage

\end{document}